\documentclass[a4paper,11pt]{article}
\pdfoutput=1 

\usepackage{jcappub} 

\usepackage[T1]{fontenc}
\usepackage{centernot}
\usepackage{tensor}
\usepackage{pgfplots}
\usepackage{xcolor}

\usepackage{graphicx,rotating,slashed,amsmath,xcolor,amssymb,amsfonts,colortbl,subfigure,float}
\usepackage{upgreek}
\usepackage{latexsym}     
\usepackage{graphics}
\usepackage{graphicx} 
\usepackage[latin1,utf8]{inputenc}
\usepackage[OT2,OT1]{fontenc}

\title{Limitations on Standard Sirens tests of gravity from screening}

\author[a]{Charles Dalang,}
\author[a]{Lucas Lombriser}

\affiliation[a]{D\'{e}partement de Physique Th\'{e}orique, Universit\'{e} de Gen\`{e}ve, \\ 24 quai Ernest Ansermet, 1211 Gen\`{e}ve 4, Switzerland}

\emailAdd{charles.dalang@unige.ch}
\emailAdd{lucas.lombriser@unige.ch}

\newcommand{\aM}{\alpha_{\rm M}}

\newcommand{\cT}{c_{\rm T}}

\def \be {\begin{equation}}
\def \ee {\end{equation}}
\def \dd {\mathrm{d}}
\def \t {\tilde}
\def \p {\partial}
\def \te {\tensor} 
\def \l {\left}
\def \r {\right}
\def \te {\tensor} 
\newcommand{\overbar}[1]{\mkern 1.5mu\overline{\mkern-1.5mu#1\mkern-1.5mu}\mkern 1.5mu}

\usepackage[normalem]{ulem}
\definecolor{darkgreen}{rgb}{0,0.5,0}
\newcommand{\edit}[1]{{\color{blue}{#1}}}

\abstract{
Modified gravity theories with an effective Newton constant that varies over cosmological timescales generally predict a different gravitational wave luminosity distance than General Relativity.
While this holds for a uniform variation, we show that if locally screened at the source and at the observer as required to pass stringent astrophysical tests of gravity, the General Relativistic distance is restored.
In the absence of such a screening, the same effect must modify electromagnetic luminosity distances inferred from supernovae Type~Ia, to the extent that the effects can cancel in the comparison.
Hence, either the modifications considered employ screening, which leaves no signature in Standard Sirens of a cosmological modification of gravity, or screening does not operate, in which case there can be a signal that is however well below the forseeable sensitivity of the probe when astrophysical bounds are employed.
We recover these results both in the Jordan and Einstein frames, paying acute attention to pecularities of each frame such as the notion of redshift or geodesic motions.
We emphasise that despite these limitations, Standard Sirens provide valuable independent tests of gravity that differ fundamentally from other probes, a circumstance that is generally important for the wider scope of gravitational modifications and related scenarios.
Finally, we use our results to show that the gravitational wave propagation is not affected by dark sector interactions, which restores a dark degeneracy between conformal and disformal couplings that enables observationally viable cosmic self-acceleration to emenate from those.
}

\begin{document}
\maketitle
\flushbottom

\section{Introduction} \label{sec:intro}

%
%
The vastly different length scales involved in cosmological tests of General Relativity (GR) compared to Solar-System probes~\cite{Will:2014kxa,Williams:2004qba} makes their execution a worthwhile endeavour~\cite{Koyama:2015vza,Joyce:2016vqv,Ishak:2018his,Ferreira:2019xrr}.
Further motivation for putting cosmological gravity under scrutiny can be drawn from the necessity of invoking an extensive dark sector in the energy budget of the Universe to explain cosmological observations in the context of GR.
Traditionally, an important driver for the development of modified theories of gravity has therefore emerged from the observed late-time accelerated expansion of our Universe~\cite{Koyama:2015vza,Joyce:2016vqv,Ishak:2018his,Ferreira:2019xrr}.
While generally expected to be attributed to a cosmological constant arising from vacuum fluctuations, it has proven extraordinarily difficult to motivate the observed acceleration from quantum theoretical predictions~\cite{Weinberg:1988cp,Martin:2012bt} (however, see, e.g., Refs.~\cite{Barrow:2010xt,Kaloper:2013zca,Wang:2017oiy,Lombriser:2018aru,Lombriser:2019jia,Alexander:2019ctv}).
Given the broad possibilities for modifying gravity, the cosmic large-scale structure was shown to be insufficient to exhaustively probe the vast available model space, being fundamentally limited by a dark degeneracy that is however broken by measurements of the cosmological propagation of gravitational waves~\cite{Lombriser:2014ira,Lombriser:2015sxa}.

The direct detection of gravitational waves from a binary black hole merger by the Laser Interferometer Gravitational-Wave Observatory (LIGO) and Virgo marked the dawn of gravitational wave astronomy~\cite{Abbott:2016blz}.
Of particular importance for cosmic tests of gravity was the simultaneous LIGO/Virgo measurement of the gravitational wave GW170817~\cite{TheLIGOScientific:2017qsa} emitted by a binary neutron star merger with a range of electromagnetic counterparts~\cite{GBM:2017lvd}.
Particularly the short gamma-ray burst following 1.7~s after the wave signal imposes a stringent constraint on the speed of gravitational waves of $|\cT^2-1|\lesssim 10^{-15}$ at redshifts of $z\lesssim0.01$~\cite{Monitor:2017mdv}, matching the forecasts of Refs.~\cite{Nishizawa:2014zna,Lombriser:2015sxa} and realising the implications for modified gravity and dark energy expressed therein (also see Ref.~\cite{Kase:2018aps} for a recent review and further references).
The measured luminal speed of gravity at late times in combination with the observed large-scale structure specifically challenges the concept that modifications of gravity could be made directly responsible for cosmic acceleration~\cite{Lombriser:2016yzn}. It is worthwhile noting that this implication could have been inferred from any of the counterpart measurement reported in Ref.~\cite{GBM:2017lvd} if a clear association with the gravitational wave event could have been established.
Caveats to the constraint on $\cT$ were expressed in Refs.~\cite{Battye:2018ssx,deRham:2018red}, pointing out that the effective field theory underlying the formulation of the broad range of modified gravity theories may break down at the high-energy scales probed by LIGO/Virgo but that future gravitational wave detectors sensitive to lower energy scales may enter the regime amenable to effective field theory.
In the upcoming years, advanced space-based detectors such as the Laser Interferometer Space Antenna (LISA) \cite{2017arXiv170200786A}, or advanced ground-based detectors such as the Einstein Telescope (ET) \cite{Punturo:2010zz}, the Kamioka Gravitational-Wave Detector (KAGRA) \cite{Somiya_2012} and LIGO-India will join the detector network and will measure gravitational waves from cosmological distances up to redshifts $z \lesssim 10$.

Besides a test of the speed of gravity, GW170817 also provides the first Standard Siren~\cite{Schutz:1986gp,Holz:2005df} from the luminosity distance measurement obtained from the decay of the wave amplitude with its travelled cosmological distance and the identification of the source redshift.
The time-dependent effective gravitational coupling of modified gravity theories affects this decay and leads to a departure of the luminosity distance measured by gravitational waves from that measured through electromagnetic means~\cite{Saltas:2014dha,Lombriser:2015sxa} (also see Refs.~\cite{Amendola:2017ovw,Nishizawa:2017nef,Belgacem:2017ihm,Belgacem:2018lbp,Linder:2018jil,Belgacem:2019pkk}). 
Forecasts for the constraints on the evolving gravitational coupling that can be inferred from Standard Sirens have, for instance, been estimated in Refs.~\cite{Lombriser:2015sxa,Belgacem:2018lbp,Belgacem:2019pkk}.
Recently, however, concerns have been expressed in Ref.~\cite{Amendola:2017ovw} on whether the modified gravitational couplings considered should be interpreted as those in the cosmological background of the emitter and observer or rather as those in their immediate environment that should be assumed screened to recover GR and comply with the stringent astrophysical tests where it has been confirmed to a high degree. 
This would therefore leave no observable effect of a cosmological gravitational modification in Standard Sirens.

In this paper we will investigate this question in detail by first providing a rigorous derivation of the luminosity distances in the Jordan frame of Horndeski scalar-tensor theory~\cite{Horndeski1974} and then rederive the equivalent expressions in its Einstein frame.
We then screen the couplings and determine the impact of screening mechanisms on the observable difference in the luminosity distances. 
Our main findings will be that the couplings that matter to Standard Sirens tests of gravity are indeed the local ones.
These are in the Jordan frame, the couplings at the source and receiver of the gravitational wave, and in the Einstein frame, the couplings entering the atomic emission lines of photons in the region of the emitter and their terrestrial laboratory counterparts, from the combination of which redshift is determined.
Hence, either local couplings are not screened, in which case one must adopt stringent astrophysical constraints on an evolving gravitational coupling such as from lunar laser ranging~\cite{Williams:2004qba}, lying several orders of magnitude beyond the sensitivity of Standard Sirens tests~\cite{Lombriser:2015sxa,Lombriser:2016yzn,Belgacem:2018lbp,Tsujikawa:2019pih,Belgacem:2019pkk} and leaving no detectable signature for those. Or the local couplings are screened, in which case the luminosity distances inferred from Standard Sirens will agree with the electromagnetic ones, as in GR.
Modified gravity effects will then be limited to a potentially different cosmological background expansion history and secondary effects in the propagation of light and gravitational waves through modified cosmological structures. 
Notably, if local couplings are assumed not screened to allow for a signature in the gravitational wave luminosity distance, the electromagnetic luminosity distance may also be modified, as we point out for the case of supernovae Type Ia in the presence of an evolving local coupling.

Nevertheless, it should be stressed that Standard Sirens provide valuable independent tests of gravity, which differ fundamentally from the more stringent astrophysical probes, a circumstance that may generally be important for more exotic gravitational modifications and related scenarios. For example, while the result likely extrapolates to more general models for which the friction in the gravitational wave propagation equation arises from a time variation of the effective Planck mass, it does not apply for some non-local gravity \cite{Belgacem:2019lwx} or extra-dimensional models. In Degenerate Higher Order Scalar-Tensor theories (DHOSTs) there are also enough free functions that may be used to get around the lunar laser ranging constraints and still generate a signal in Standard Sirens~\cite{Crisostomi:2017lbg,Belgacem:2019pkk}.

Employing our Einstein-frame calculations, we furthermore explore the gravitational wave propagation for non-universal conformal and disformal couplings to a scalar field.
In particular we examine dark sector interactions, where baryons and photons remain minimally coupled, and show that the gravitational waves are unaffected by the nonminimal couplings in the dark sector.
We discuss how due to the recovery of the dark degeneracy~\cite{Lombriser:2014ira,Lombriser:2015sxa}, lacking the tensor speed constraint, such interactions can yield an observationally compatible self-acceleration effect emanating from the dark sector couplings.
In analogy to Ref.~\cite{Lombriser:2016yzn} we compute the minimal threshold for completely attributing self-acceleration to the evolution of conformal and disformal scalar dark sector couplings.

The paper is organized as follows.
In Sec.~\ref{sec:standardsirenstests}, we briefly review various aspects of Horndeski theory relevant to our analysis.
In Sec.~\ref{sec:JordanFrame}, we present the derivations of the electromagnetic and gravitational wave luminosity distances for the subset of Horndeski gravity models that respect the luminal speed of gravitational waves.
We then discuss the luminosity distances in the Einstein frame in Sec.~\ref{sec:EinsteinFrame}, examining a number of subtleties that enter the computation such as the correct notion of redshift and the geodesic motion.
In Sec.~\ref{sec:screeningmechanism}, we clarify how screening of the effective gravitational coupling affects the luminosity distances and redshift both in the Jordan and Einstein frames.
Employing our results from the Einstein frame in Sec.~\ref{sec:darksectorinteractions}, we consider non-universal couplings of the scalar field to matter, where baryonic and dark matter particles couple minimally to different metrics, related by conformal or disformal transformations.
We also discuss here how an observationally viable cosmic self-acceleration effect may originate from the dark sector interactions due to the dark degeneracy between conformal and disformal couplings in the large-scale structure.
Finally, we conclude in Sec.~\ref{sec:conclusions}.

\section{Standard Sirens tests of gravity} \label{sec:standardsirenstests}

Before engaging in the detailed calculations of the observable effects of alternative gravity theories on Standard Sirens in Secs.~\ref{sec:JordanFrame}--\ref{sec:screeningmechanism}, we shall start with a short introduction to Horndeski scalar-tensor theories in Sec.~\ref{sec:horndeski}.
In Sec.~\ref{sec:standardsirens} we will briefly discuss the main concepts behind the Standard Sirens tests of gravity, but we refer the reader to Secs.~\ref{sec:JordanFrame}--\ref{sec:screeningmechanism} for a more rigorous discussion of the various physical aspects involved.

\subsection{Horndeski scalar-tensor modifications of gravity and screening mechanisms} \label{sec:horndeski}

We will focus our analysis of Standard Sirens tests of gravity on the Horndeski action~\cite{Horndeski1974,Deffayet:2011gz,Kobayashi:2011nu} (see Ref.~\cite{Belgacem:2019pkk} for a broader discussion), which describes the most general, four-dimensional, Lorentz-covariant scalar-tensor theory that produces at most second-order equations of motion.
However, it will become clear that the same conclusions that will be inferred may more generally apply to theories that exhibit a slowly varying effective Planck mass.
The Horndeski action is given by
\begin{equation}
 S = \int \dd^4x \sqrt{-g} \left[ \frac{M_p^2}{2} \sum_{i=2}^5 \mathcal{L}_i + \mathcal{L}_{\rm m}(g_{\mu\nu},\psi_{\rm m}) \right] \,, \label{eq:horndeski}
\end{equation}\label{Horndeski_Action}
where the different Lagrangian densities are specified by
\begin{eqnarray}
 \mathcal{L}_2 & = & G_2(\phi,X) \,, \\
 \mathcal{L}_3 & = & G_3(\phi,X)\Box \phi \,, \\
 \mathcal{L}_4 & = & G_4(\phi,X)R + G_{4X}(\phi,X) \left[ (\Box\phi)^2 - (\nabla_{\mu}\nabla_{\nu}\phi)^2 \right] \,, \\
 \mathcal{L}_5 & = & G_5(\phi,X) G_{\mu\nu}\nabla^{\mu}\nabla^{\nu}\phi - \frac{1}{6} G_{5X}(\phi,X) \left[ (\Box\phi)^3 - 3\Box\phi(\nabla_{\mu}\nabla_{\nu}\phi)^2 + 2 (\nabla_{\mu}\nabla_{\nu}\phi)^3 \right]
\end{eqnarray}
with $X\equiv-\frac{1}{2}\partial_{\mu}\phi\partial^{\mu}\phi$ and minimally coupled matter fields $\psi_{\rm m}$ in $\mathcal{L}_{\rm m}$.
$R$ and $G_{\mu\nu}$ denote the Ricci scalar and Einstein tensor of the Jordan frame metric $g_{\mu\nu}$, and $M_p$ is the bare Planck mass.
The constraint on the speed of gravitational waves implies that~\cite{Kimura:2011qn,McManus:2016kxu}
\be
G_{4X}=G_5\simeq 0 \,, \label{eq:Gconstraints}
\ee
in the low-redshift regime (see, however, Refs.~\cite{Battye:2018ssx,deRham:2018red,Copeland:2018yuh}).
This considerably simplifies the action~\eqref{Horndeski_Action}, and we shall adopt this constraint in the following discussion, except for when examining dark sector interactions in Sec.~\ref{sec:darksectorinteractions}, which are not subdued to this bound.
The remaining nonminimal coupling $G_4(\phi)$ to the Ricci scalar yields a spacetime dependent effective Planck mass
\begin{equation}
M(t,\mathbf{x}) = M_p \sqrt{G_4(\phi(t,\mathbf{x}))} \,.
\end{equation}
and correspondingly an effective Newton constant $G_{\tiny{\hbox{eff}}}(t,\mathbf{x}) = G_N /G_4(\phi(t,\mathbf{x}))$.

Due to stringent astrophysical constraints on departures from GR~\cite{Will:2014kxa,Williams:2004qba}, observationally viable Horndeski theories must employ screening mechanism that recover GR in high-density regions such as the Solar System.
The different classes of screening mechanisms~\cite{Joyce:2016vqv} can be divided into those that screen the equations of motion through large potential wells, $\Phi_N>\Lambda$, for some threshold value $\Lambda$ such as for the chameleon effect~\cite{Khoury:2003aq}; through large first derivatives, $\nabla\Phi_N>\Lambda^2$, such as in k-mouflage models~\cite{Babichev:2009ee}; and through large second derivatives, $\nabla^2\Phi_N>\Lambda^3$, such as for the Vainshtein mechanism~\cite{Vainshtein:1972sx}. Also see Refs.~\cite{Brax:2019koq,Lombriser:2014ira} for further shielding mechanisms. While the chameleon mechanism recovers $G_{\tiny{\hbox{eff}}}=G_N$ in a screened region, the background evolution of $G_{\tiny{\hbox{eff}}}$ may not be screened by the Vainshtein mechanism~\cite{Babichev:2011iz}.
It is worth noting that lunar laser-ranging constraints~\cite{Williams:2004qba} would therefore in principle already have ruled out evolving $G_4$ and $G_5$ for shift-symmetric theories such as Galileon gravity~\cite{Babichev:2011iz} years before the gravitational wave speed constraint in Eq.~\eqref{eq:Gconstraints} (also see Refs.~\cite{Jimenez:2015bwa,Lombriser:2015sxa}).
Importantly, however, this conclusion cannot be generalised to the full Horndeski action~\cite{Lombriser:2015sxa} and the extrapolation of $G_{\tiny{\hbox{eff}}}$ in the highly nonlinear Earth-Moon system to the effective coupling in the cosmological background should be taken as a strong caveat that may involve complications that would not allow a straightforward connection. Moreover, one may question the applicability of effective field theory for the lunar laser-ranging experiments (cf.~\cite{deRham:2018red}).
Here we will assume that a viable effective screening mechanism should imply the screening of the evolution of the Planck mass in high-density regions, and we will revisit the effect of screening in Sec.~\ref{sec:screeningmechanism}.
Finally, it should be noted that lunar laser ranging does not directly test $G_{\tiny{\hbox{eff}}}$ but the effective gravitational coupling that modifies the Earth-Moon dynamics~\cite{Sakstein:2017pqi}.
However, together with stringent Shapiro time delay constraints in the Solar System on the Eddington parameter $\gamma_{\rm PPN}$ this translates into a constraint on $G_{\tiny{\hbox{eff}}}$, where a degeneracy in the modifications is excluded due to the constraint on the speed of gravitational waves~\cite{Lombriser:2015sxa}.

For further details on Horndeski theory, the variety of observational constraints on it, and the screening mechanisms employed, we refer to the reviews in Refs.~\cite{Koyama:2015vza,Joyce:2016vqv,Ishak:2018his,Ferreira:2019xrr}.
In the following, we focus on one particular test of scalar-tensor theories that exploits the effect of an evolving Planck mass on the cosmological propagation of gravitational waves.

\subsection{Standard Sirens} \label{sec:standardsirens}

Besides the speed of gravitational waves tested by comparing the arrival times of the wave to electromagnetic counterparts~\cite{Nishizawa:2014zna,Lombriser:2015sxa,Monitor:2017mdv,GBM:2017lvd}, another observable is the luminosity distance $d_L^{gw}$ inferred from the decay of the wave amplitude with its travelled distance~\cite{Schutz:1986gp,Holz:2005df,Saltas:2014dha,Lombriser:2015sxa,Amendola:2017ovw,Nishizawa:2017nef,Belgacem:2017ihm,Belgacem:2018lbp,Linder:2018jil,Belgacem:2019pkk} (see Sec.~\ref{sec:JordanFrame} for the details).
An additional identification of the redshift of the gravitational wave source provides a distance-redshift relation, prompting the use of Standard Sirens~\cite{Holz:2005df} for reference to the qualifying binary mergers.
In GR, this distance agrees with that inferred from the electromagnetic luminosity distance $d_L^{em}$, obtained for example from the flux measurement of supernovae Type Ia~\cite{Maggiore:1900zz}.
However, in modified theories of gravity which exhibit a time-varying effective gravitational constant $G_{\tiny{\hbox{eff}}}(t)$, the friction term entering the tensor wave equation differs from that of photons, leading to distinct damping effects on the amplitudes of light and of gravitational waves, thus causing a relative departure between the respective cosmological luminosity distances inferred ($d_L^{gw} \neq d_L^{em}$)~\cite{Saltas:2014dha,Lombriser:2015sxa}. The ratio of the luminosity distances can be shown to be simply given by the ratio of the effective Planck mass at the observer to that at the time of emission~\cite{Amendola:2017ovw,Belgacem:2017ihm,Linder:2018jil,Belgacem:2019pkk}
\be
 \frac{d_L^{gw}(z)}{d_L^{em}(z)}= \frac{M(0)}{M(z)} \label{Ratio}
\ee
with the source at redshift $z$ (we shall rederive this expression in Sec.~\ref{sec:JordanFrame}, however, also consider Eq.~\eqref{eq:DLgwtoDLSNIa} for a deviating expression related to peculiarties in the electromagnetic source).
Accurate measurements of gravitational wave and electromagnetic luminosity distances therefore provide a constraint on the evolution of the effective gravitational coupling. 
A first forecast for the bounds that can be placed on this evolution by Standard Sirens tests was inferred in Ref.~\cite{Lombriser:2015sxa} for LISA.
A more elaborate, updated analysis for LISA forecasts was conducted more recently in Ref.~\cite{Belgacem:2019pkk}. Similarly, \edit{Refs.}~\cite{Belgacem:2018lbp,Lagos:2019kds,Nunes:2019bjq} estimated constraints on the ratio~\eqref{Ratio} achievable with \edit{LIGO or} the Einstein Telescope.
Further forecasts for the Einstein Telescope, Cosmic Explorer, or Voyager were made in Ref.~\cite{Nishizawa:2019rra} and Ref.~\cite{Arai:2017hxj} inferred constraints on the evolution of the Planck mass from GW170817.
\begin{figure}
\begin{center}
\begin{tikzpicture}
\draw [->] (0,0.5) -- (9.8,0.5);
\draw [->] (0,0.5) -- (0,6.3);
\draw (10,0.5) node[right,scale=0.8] {$r$};
\draw (9.2,0.4) -- (9.2,0.6);
\draw (0,0.4) -- (0,0.6);
\draw (9.2,0.4) node[below, scale=0.8] {$r_o$};
\draw (0,0.4) node[below, scale=0.8] {$r_s$};
\draw (0,6.3) node[above, scale=0.8] {$M^2(r(t))$};
\draw (0,1.5) node[left, scale=0.8] {$M_p^2$};
\draw[fill=black!13!blue!7]
      [dashed](8.15,0.5)--(8.15,5.8) -- (5.1,5.8) -- (5.1,0.5);
\draw[fill=black!13!blue!13]
      [dashed](8.15,0.5)--(8.15,5.8) -- (9.2,5.8) -- (9.2,0.5);
\draw (8.65,3.7) node[scale=0.8, rotate=90] {screened observer};
\draw[fill=black!13!blue!13]
      [dashed](0,0.5)--(0,5.8) -- (1.05,5.8) -- (1.05,0.5);
\draw[fill=black!13!blue!7]
      [dashed](1.05,0.5)--(1.05,5.8) -- (2.1,5.8) -- (2.1,0.5);
\draw[fill=black!13!blue!7]
      [dashed](1.05,0.5)--(1.05,5.8) -- (2.1,5.8) -- (2.1,0.5);
\draw (0.525,3.7) node[scale=0.8, rotate=90] {screened emitter};
\draw (1.53,3.7) node[scale=0.8, rotate=90] {transition zone};
\draw[fill=black!18!blue!2]
      [dashed](2.1,0.5)--(2.1,5.8) -- (5.5,5.8) -- (5.5,0.5);
\draw (3.0,3.2) node[above, scale=0.7] {$M'/M >0$};
\draw (7.35,3.75) node[above,scale=0.7] {$M'/M<0$};
\draw (3.8,1.2) node[below,scale=0.8] {background evolution};
\draw (5.6,1) node[right,scale=0.8] {transition zone};
\draw [dashed] (0,1.5) -- (9.2,1.5);
  \draw plot coordinates {(0, 1.5) (1/5, 1.5*1.001) (2/5,1.5* 1.003) (3/5, 1.5*1.011) (4/5, 1.5*1.017) (5/5, 1.5*1.032) (6/5,1.5* 1.05) (7/5, 1.5*1.08) (8/5, 1.5*1.15) (9/5, 1.5*1.25) (10/5, 1.5*1.35) (11/5, 1.5*1.45) (12/5, 1.5*1.55) (13/5, 1.5*1.65) (14/5, 1.5*1.75) (15/5, 1.5*1.85) (16/5, 1.5*1.95) (17/5, 1.5*2.05) (18/5, 1.5*2.15) (19/5, 1.5*2.25) (20/5, 1.5* 2.35) (21/5, 1.5*2.45) (22/5, 1.5*2.55) (23/5, 1.5*2.65) (24/5, 1.5*2.75) (25/5, 1.5*2.85) (26/5, 1.5*2.95) (27/5, 1.5*3.05) (28/5,1.5*3.10) (29/5, 1.5* 3.12) (30/5, 1.5*3.11) (31/5, 1.5*3.07) (31.5/5, 1.5*3.02) (32/5, 1.5*2.9) (32.25/5, 1.5*2.85) (32.5/5, 1.5*2.79)(33/5, 1.5*2.6) (34/5,1.5*2) (35/5, 1.5*1.5) (36/5, 1.5*1.19) (36.5/5, 1.5*1.13) (37/5, 1.5*1.09) (38/5, 1.5*1.06) (39/5,1.5*1.04) (40/5, 1.5*1.032) (41/5, 1.5*1.017) (42/5, 1.5*1.011) (43/5, 1.5*1.003) (44/5,1.5*1.001) (45/5, 1.5*1.0005) (46/5, 1.5)};
\end{tikzpicture}
\caption{
A gravitational wave experiences a friction $M^{-1}(r(t)) \dd M(r(t))/ \dd t$ from the evolution of the effective Planck mass $M$.
At the screened source and observer, the bare Planck mass must be restored such that $M \to M_p$.
The gravitational wave progressively leaves the screened region and enters a transition to a propagation through the average cosmological medium.
It experiences a positive friction $M'(r(t))/M(r(t))>0$ due to the background evolution until it reaches the observer, where it transitions back into a screened region.
The total enhancement in $M$ accumulated during the propagation to the transition region of the observer must now be compensated by an equivalent suppression by the negative friction $M'(r(t))/M(r(t))<0$ during the transition to the screened region of the observer such that $M=M_p$ is recovered.
The net effect of the varying friction on the gravitational wave amplitude $\propto M$ and hence the luminosity distance therefore cancels out exactly once the wave reaches the screened region of the observer.
}
\label{fig:Planck Mass Evolution}
\end{center}
\end{figure}
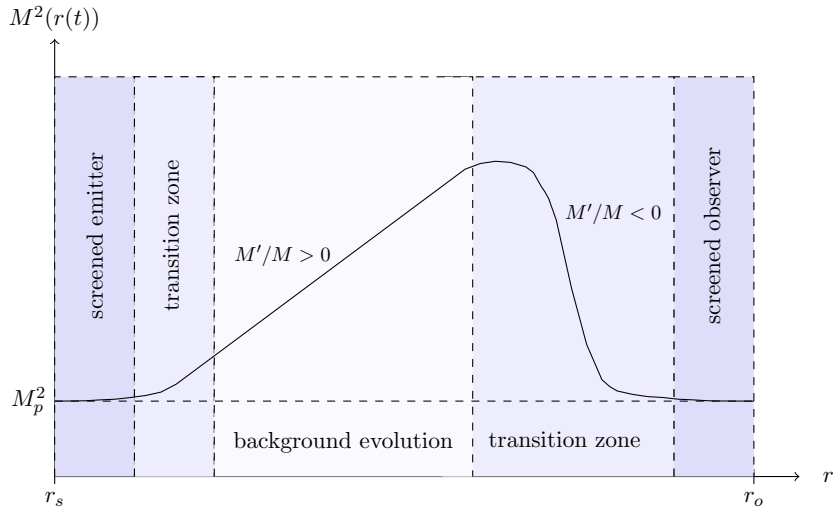
Some concerns have recently been raised in Ref.~\cite{Amendola:2017ovw} on whether the Planck masses in Eq.~\eqref{Ratio} should be interpreted as those in the cosmological background of the emitter and observer or rather as those in their immediate environment, which hence should be assumed screened to comply with stringent astrophysical tests~\cite{Will:2014kxa,Williams:2004qba}, recovering the GR value $M=M_p$ in Eq.~\eqref{Ratio} and leaving no observable effect of the gravitational modification in the Standard Sirens. 

We will investigate this question in Secs.~\ref{sec:JordanFrame}--\ref{sec:screeningmechanism} by first providing a rigorous derivation of Eq.~\eqref{Ratio} in the Jordan frame of action~\eqref{Horndeski_Action} (Sec.~\ref{sec:JordanFrame}) and then rederive the equivalent expressions in the Einstein frame (Sec.~\ref{sec:EinsteinFrame}) to then screen the nonminimal couplings and determine the impact of screening effects on Eq.~\eqref{Ratio} (Sec.~\ref{sec:screeningmechanism}). 
We advise the reader familiar with Standard Sirens tests of gravity to skip Sec.~\ref{sec:JordanFrame} and directly proceed to Sec.~\ref{sec:EinsteinFrame}.

As alluded to in the introduction, we will find that the couplings that matter to Standard Sirens tests are indeed the local ones, and hence that Standard Sirens can be considered screened for observational compatibility with astrophysical probes.
An intuitive understanding of the phenomenon in the Jordan frame is presented in Fig.~\ref{fig:Planck Mass Evolution}.
The Planck mass is screened at the source $M=M_p$ and later gets modified outside the screened region and during its propagation to the observer. As the gravitational wave reaches the detectors of the observer in the Solar System, the Planck mass recovers $M= M_p$.
As the damping of the gravitational wave depends on the evolution rate of $M$ (see Eq.~\eqref{alphaM}), the suppression of the wave amplitude will be compensated by an equivalent enhancement before reaching the detector and hence will leave no signature of the gravitational modification in the Standard Siren.

Furthermore, if replacing the depicted Planck mass evolution in Fig.~\ref{fig:Planck Mass Evolution} by an evolution of the gravitational wave speed $\cT$ it also becomes apparent for why in contrast we do not expect such cancellations to happen for the observed arrival time difference between gravitational waves and light. The latter may na\"ively be written as 
\be
\dd t = \dd r \l(\frac{1}{c}- \frac{1}{\cT(r)} \r),
\ee
which leads to
\be
\Delta t= \frac{(r_o - r_s)}{c} - \int_{r_s}^{r_o} \frac{\dd r}{\cT(r)} = \mathcal{O}(1 \hbox{ second}) \,.
\ee
Hence the delay of one signal over the other, as can be illustrated analogously to Fig.~\ref{fig:Planck Mass Evolution}, only accumulates with no further delay added where the signals propagate through a screened region.
A chance cancellation of an enhancement of $\cT$ with an equivalent suppression at the level of $\mathcal{O}(1 \hbox{ second})$ for a travel time of $\mathcal{O}(10^8 \hbox{ years})$ seems unreasonably coincidental.

\section{Light and gravitational wave propagation in the Jordan frame}\label{sec:JordanFrame}

We shall now perform a more rigorous computation of the effect of modified gravity on the ratio of gravitational wave and electromagnetic luminosity distances in the absence of a screening of the local Planck mass evolution, first adopting the Jordan frame. 
We start with a discussion of electromagnetic radiation and the impact of an expanding universe in Sec.~\ref{sec:EMradiation}.
Considering the geometric optics approximation, we compute the flux and derive $d_L^{em}$ in Sec.~\ref{sec:optics}. We also discuss the production mechanism of supernovae Type Ia.
Finally, in Sec.~\ref{sec:JFGWs} we calculate the gravitational wave luminosity distance $d_L^{gw}$ and compare it to its electromagnetic counterpart.


\subsection{Electromagnetic radiation in an expanding universe} \label{sec:EMradiation}

Consider the electromagnetic action in curved spacetime
\be
S_{em} = -\frac{1}{4 \pi}  \int \mathrm{d}^4x \sqrt{-g} F_{\mu \nu} F^{\mu \nu} \,, \label{eq:EMaction}
\ee
where $A_\mu$ denotes the electromagnetic potential and $F_{\mu\nu} \equiv \nabla_\mu A_\nu - \nabla_\nu A_\mu$, the field strength tensor. Extremisation with respect to $A_{\nu}$ yields the Euler-Lagrange equations 
\be
\frac{\p\mathcal{L}_{em}}{\p A_\nu} - \nabla_\mu \frac{\p \mathcal{L}_{em}}{\p (\nabla_\mu A_\nu)} =0
\ee
such that
\be 
 \nabla_\mu (\nabla^\mu A^\nu - \nabla^\nu A^\mu) = \nabla_\mu \nabla^{\mu} A^\nu - \nabla^\nu \nabla_\mu A^\mu  - \tensor{R}{^\nu_\mu} A^\mu = 0 \,, \label{eq:curvedPhotonEoM}
\ee
where the Ricci curvature is introduced by the commutation of the covariant derivatives. Varying Eq.~\eqref{eq:EMaction} with respect to $A^\nu$ instead yields
\be
 \nabla^\mu F_{\mu \nu}=0 \,. \label{eq:F1}
\ee
Imposing the curved spacetime generalisation of the Lorenz gauge, $\nabla_\mu A^\mu = 0$, still allows for some gauge freedom.
This is because the four-vector $A^\mu$ has four components but only two propagating degrees of freedom. To fix the gauge we impose the additional condition $A_0=0$ and we arrive at the photon wave equation in curved spacetime
\be\label{curvedPhotonEoM}
\nabla_\mu \nabla^{\mu} A^\nu  - \tensor{R}{^\nu_\mu} A^\mu = 0 \,.
\ee
The Ricci tensor
acts as a mass term for the field $A^\mu$.
In the limit where this term can be neglected, which is known as the geometric optics or eikonal approximation (Sec.~\ref{sec:optics}), photons can be shown to follow null geodesics.

We assume a flat statistically spatially homogeneous and isotropic universe, where the Friedmann-Lema\^itre-Robertson-Walker (FLRW) metric is defined by the line element
$\dd s^2 =  a^2(\eta) \left(-\dd \eta^2 + \dd r^2 + r^2 \dd\Omega^2 \right)$
with scale factor $a$, conformal time $\eta$, comoving coordinate distance $r$, and volume element $\dd\Omega^2$. Eqs.~\eqref{eq:F1} and \eqref{eq:curvedPhotonEoM} become
\begin{align}
 \begin{cases} \label{eq:JframeA2}
  \partial_\eta^2 A_\sigma -\vec{\nabla}^2 A_\sigma = 0 \,, \\ 
  \partial_\eta^2 A^\nu + 4 \mathcal{H} \partial_\eta A^\nu -\l(\vec{\nabla}^2 -2 \frac{a'' a + (a')^2}{a^2}\r)A^\nu = 0 \,,
 \end{cases}
\end{align}
respectively, where primes denotes derivatives with respect to $\eta$, here and throughout the article, unless otherwise specified and $\mathcal{H} \equiv a'/a$ is the Hubble function. Importantly, note that the equation of motion for $A^\nu$ is different from that of $A_\sigma$, which leads to different damping effects on cosmological scales.

We now want to find the solutions to Eqs.~\eqref{eq:JframeA2}.
Since the waves propagate over a two-sphere, we use spherical coordinates and define the functions $B_\sigma$ and $B^\nu$ such that $A(t,r,\theta,\phi) = \frac{1}{r} B(t,r)$.
It follows that
\begin{align}
 \begin{cases}
  \partial_\eta^2 B_\sigma -\partial_r^2 B_\sigma = 0 \,,\\
  \partial_\eta^2 B^\nu + 4 \mathcal{H} \partial_\eta B^\nu- \l(\p_r^2 -2\frac{a'' a + (a')^2}{a^2} \r) B^\nu = 0 \,.
 \end{cases}
\end{align}
Performing Fourier transformations we obtain 
\begin{align}
 \begin{cases}
  \partial_\eta^2 B_\sigma +k^2 B_\sigma = 0 \,,\\
  \partial_\eta^2 B^\nu + 4 \mathcal{H} \partial_\eta B^\nu+ \l(k^2 + 2\frac{a'' a + (a')^2}{a^2}\r)B^\nu = 0 \,,
 \end{cases}
\end{align}
where for convenience we do not introduce any notation to indicate transformed quantities.
The first equation is that for a standard wave.
Recall the $1/r$ damping.
Hence,
\be
 A_\sigma(\eta, r)= \sum_{\lambda=\pm}\frac{A_1^{(\lambda)}}{r a_s} \cos \left( k r - \omega \eta + \varphi_1^{(\lambda)} \right) \epsilon_\sigma^{(\lambda)} \,,
\ee
where $A_1^{(\lambda)}$, $\varphi_1^{(\lambda)}$ are integration constants specified by the initial conditions which may depend on one of the two physical polarisation states parametrized by $\lambda$, $\epsilon_\sigma^{(\lambda)}$ is a polarisation vector and we introduced the scale factor at the source of the wave $a_s\equiv a(t_s)$ at the time of emission $t_s$ in the denominator to transform the coordinate distance $r$ into a physical distance $r_{phys} = a_s r$.
Note that $r = \int \dd \eta$ and $r_{phys} =\int \dd s = \int a(t_s) \dd r = a(t_s)\int \dd r = a(t_s) r $ represents a physical distance at a given cosmic time, here $t_s$ ($\dd t =0$).
To solve for $B^\nu$ we define $B^\nu(\eta, k) \equiv a^{-2} \chi^\nu (\eta, k)$ such that for $\chi$ one gets the standard wave equation $\p_\eta^2 \chi^\nu (\eta,k) + k^2 \chi^\nu (\eta, k)=0$.
Therefore,
\be
 A^\nu(\eta, r) = \sum_{\lambda=\pm} \frac{a_s^2 A_{2}^{(\lambda)}}{a^2(\eta) r a_s } \cos \l( k r - \omega \eta + \varphi_{2}^{(\lambda)}\r) \epsilon^\nu_{(\lambda)}
\ee
with polarisation vector $\epsilon^{\nu}_{(\lambda)}$ and $A_2^{(\lambda)}$, $\varphi_2^{(\lambda)}$ specified by the initial conditions.
In order to preserve the relationship $A^\nu = g^{\nu \sigma} A_\sigma$ in conformal time and because we only measure products of $A_\sigma$ with $A^\nu$, we can furthermore redefine the integration constants $\varphi_I^{(\lambda)}\equiv \varphi_1^{(\lambda)}=\varphi_2^{(\lambda)}$ and $ A_I^{(\lambda)} = a_s^{-1}A_1^{(\lambda)} = a_s A_2^{(\lambda)}$, such that
\begin{align}
A_\sigma(\eta, r)= &\sum_{\lambda= \pm}\frac{A_I^{(\lambda)} a_s}{r a_s} \cos \left( k r - \omega \eta + \varphi_I^{(\lambda)} \right) \epsilon_\sigma^{(\lambda)}\\ A^\nu(\eta, r) = & \sum_{\lambda=\pm}\frac{a_s}{a^2(\eta)} \frac{A_I^{(\lambda)}}{r a_s} \cos \l( k r - \omega \eta + \varphi_I^{(\lambda)} \r)\epsilon^\nu_{(\lambda)} \,. \label{eq:JFsolutions}
\end{align}
Importantly, the cosmological propagation damps $A^\nu$ with an extra factor $a^{-2}(\eta)$ with respect to $A_\sigma$.
Of course, we recognise here the general relationship $A^\nu = g^{\nu \sigma}A_\sigma$.

\subsection{Optics and flux} \label{sec:optics}

In order to compute the measured photon flux at the observer, we shall adopt and briefly review the geometric optics approximation.
We refer the reader to Ref.~\cite{Peter:2013avv} for details.
We make here the ansatz $A_\mu = C_\mu e^{i \varphi}$ for the four-vector potential and assume a slow variation of the amplitude with respect to the phase.
This approximation is well justified given the cosmological decay of the amplitudes over megaparsec scales and that the photon wavelengths are smaller than a few meters, or kilometers for gravitational waves. Schematically, we have $\p C_\mu \ll \p \varphi$. The Lorenz gauge condition and the equation of motion for $A_\sigma$ in Eq.~\eqref{eq:JframeA2} imply
\be
 \p^\mu \p_\mu \varphi = 0 \quad \textrm{and} \quad \p_\mu \varphi \p^\mu \varphi =0 \,. \label{eq:nullgeodesics}
\ee
We define the photon wavevector $k^\mu \equiv \p^\mu \varphi$, which is orthogonal to the surface on which $\varphi = cst$.
The second relation in Eq.~\eqref{eq:nullgeodesics} is of course known as the null geodesic equation $k_\mu k^\mu =0$.
Differentiation and rearranging yields the geodesic equation $k^\mu \nabla_\mu k^\nu =0$.
We may rewrite $k^\mu = \frac{\mathrm{d}x^\mu}{\mathrm{d}\lambda}$, where $\lambda$ is an affine parameter, which yields the perhaps more familiar expression
\be
\frac{\mathrm{d}k^\nu}{\mathrm{d}\lambda} + \Gamma^\nu_{\alpha\beta}k^\alpha k^\beta =0 \,.
\ee
From this, it is straightforward to show that
\be
\frac{\mathrm{d} k_\mu}{\mathrm{d} \lambda} = \frac{1}{2} \l( g_{\alpha \kappa, \mu}\r) k^\alpha k^\kappa \,.
\ee
Since the FLRW metric in conformal time can be written as $g_{\mu\nu} = a^2(\eta)\eta_{\mu\nu}$ with Minkowski metric $\eta_{\mu\nu}$, we have $g_{\mu\nu, i} =0$, which implies $k_i = cst$ along geodesics and $k^i = a^{-2}k_i$.
The null geodesic equation~\eqref{eq:nullgeodesics} ($k_\mu k^\mu =0$) implies $k_0 = cst$ and $k^0 = a^{-2}k_0$.
In light of the geometric optics approximation and of the Jordan frame solutions in Sec.~\ref{sec:EMradiation}, we may identify $C_\mu$ and $\varphi$ in Eq.~\eqref{eq:JFsolutions} from
\be
Re (C_\mu e^{i \varphi}) = C_\mu \cos(\varphi)= \frac{A_I^{(\lambda)}}{r a_s} \epsilon_\mu^{(\lambda)} \cos(kr - \omega \eta + \varphi_I^{(\lambda)}) \,,
\ee
and hence,
\be
 C_\mu = \frac{A_I^{(\lambda)}}{r a_s} \epsilon_\mu^{(\lambda)} \,, \quad \varphi = kr - \omega \eta + \varphi_I^{(\lambda)}= \omega(r- \eta) + \varphi_I^{(\lambda)} \,,
\ee
where $k = \omega$. Therefore,
\be
 \p_\mu A_\sigma = (\p_\mu C_\sigma) \cos(\varphi) - C_\sigma (\p_\mu \varphi) \sin(\varphi) \simeq  - C_\sigma k_\mu \sin(\varphi) \,.
\ee
Taking derivatives of $A_\sigma$ thus yields a $k_\mu$ factor that is constant along null geodesics.
This becomes useful when computing the photon flux at the observer.
The measurement of the flux of photons $F$ is what determines the electromagnetic luminosity distance $d_L^{em}$ to an object. More specifically,
\be
 F = \frac{\mathcal{L}_s}{4 \pi (d_L^{em})^2} \,,
\ee
where $\mathcal{L}_s$ is the luminosity in the source frame in units of energy per unit time. Suppose that we wish to measure the distance to a galaxy that is located in the z-direction.
We may use a standard candle of calibrated luminosity, measure the flux, and infer $d_L^{em}$.
We stress however that supernovae Type Ia are not standard candles in the case where the time variation of $G_{\tiny{\hbox{eff}}}$ is not screened as explained below Eq. \eqref{eq:JordanFrameL}. For the following, we simply assume that standard candles with fixed calibrated luminosity $\mathcal{L}_s$ exist.

We will assume that the observer and the source are at rest in comoving coordinates and express the four-velocity of the observer in conformal time as
\be
 u^\mu = \frac{\dd x^\mu}{\dd \tau} =  \frac{\dd \eta}{\dd \tau}\frac{\dd x^\mu}{\dd \eta} = a^{-1} \l(\frac{\dd \eta}{\dd \eta},\frac{\dd x}{\dd \eta} ,\frac{\dd y}{\dd \eta},\frac{\dd z}{\dd \eta} \r) = (a^{-1},0,0,0) \,.
\ee
It follows from the geodesic equations that the wavevector of an emitted photon is given by $k^\mu=a^{-2}\omega{\l(1, 0,0,1\r)}$ with $\omega = cst$.
The angular frequencies $\omega_s$ and $\omega_o$ of the photon emitted at the source and measured at the observer, respectively, are given by
\begin{align}
 -\omega_s &\equiv (k \cdot u )_s = g_{\mu\nu}(S) k^\mu (S) u^\nu(S) = -a_s^{-1} \omega \,,\\
 -\omega_o &\equiv (k \cdot u )_o= g_{\mu\nu}(O) k^\mu (O) u^\nu(O) = - a_o^{-1} \omega \,,
\end{align}
which defines the redshift as
\be
 1 + z \equiv \frac{(k \cdot u )_s}{ (k \cdot u )_o} = \frac{\omega_s}{\omega_o} = \frac{a_o}{a_s}
\ee
with $a_o$ denoting the scale factor at the observer (typically set to unity).

Given the electromagnetic action~\eqref{eq:EMaction} in Sec.~\ref{sec:EMradiation}, we may define the corresponding energy-momentum tensor by
\be
 T_{\mu\nu}^{em} = \frac{1}{\pi} \l( \te{F}{_\mu ^\beta} F_{\nu \beta} - \frac{1}{4} g_{\mu \nu} F_{\alpha \beta}F^{\alpha \beta }\r) \,, \label{eq:EMTmunu}
\ee
which follows from
\be
T_{\mu\nu}^{em} \equiv -\frac{2}{\sqrt{-g}} \frac{\p \l( \mathcal{L}_{em} \sqrt{-g} \r)}{\p g^{\mu \nu}} = -2 \frac{\p \mathcal{L}_{em}}{\p g^{\mu \nu}} + g_{\mu \nu} \mathcal{L}_{em}
\ee
for the Lagrangian density $4 \pi\mathcal{L}_{em} = - F_{\mu \nu} F^{\mu \nu} = - g^{\mu \alpha} g^{\nu \beta} F_{\mu \nu} F_{\alpha \beta}$. This allows one to formulate more generally the flux of the electromagnetic field measured in the direction $n^\alpha$ as~\cite{Peter:2013avv}
\be
F= - T_{\mu\nu}^{em} u^\mu \gamma^{\nu \alpha} n_\alpha
\ee
with $\gamma^{\nu\alpha} \equiv g^{\nu\alpha} + u^\nu u^\alpha$.
Choosing $n^\alpha$ such that $u^\alpha n_\alpha=0 $ one gets
\be
 F= - T_{\mu\nu}^{em} u^\mu n^\nu \,.
\ee
The detector shall lie in the $x$-$y$ plane whereas the emitted light arrives from the $z$-direction.
The four-vectors $n_o$ and $n_s$ are normalised spacelike vectors pointing into the photon direction at the observer and at the source respectively, hence,
\begin{align}
 n_o = & \frac{1}{\omega_o}(k^\mu_o + (k \cdot u)_o u_o^\mu) \,,\\
 n_s = & \frac{1}{\omega_s}(k^\mu_s + (k \cdot u)_s u_s^\mu) \,.
\end{align}
In our setting, $n_o^\nu = (0,0,0,a_o^{-1})$ and $n_s^\nu= (0,0,0,a_s^{-1})$. 
To compute the flux, we will calculate the sum over polarisations using the fact that the polarisation vectors form an orthonormal basis of the plane orthogonal to the photon wave vector
\be
\sum_{\lambda=\pm} \sum_{\lambda'=\pm}  \epsilon_\alpha^{(\lambda)} \epsilon_{(\lambda')}^\alpha = \sum_{\lambda=\pm} \sum_{\lambda'= \pm} g_{\alpha \beta} \epsilon^\alpha_{(\lambda)} \epsilon_{(\lambda')}^\beta = \sum_{\lambda=\pm} \sum_{\lambda'=\pm} \delta_{\lambda \lambda'} =2\,.
\ee

We can now compute the invariant flux in the rest frame of the observer, assumed at rest with respect to the Hubble flow. 
The observed flux $F_o$ is determined by
\begin{align} \label{eq:JordanFlux}
F_o = & -\l(T_{\mu\nu}^{em} u^\mu n^\nu \r)_o = -\l(T_{0z}^{em} n^z\r)_o u_o^0 
= a_o^{-1} n_o^z \l( \frac{1}{\pi} g^{\alpha \beta} F_{0\alpha} F_{z \beta} - \underbrace{g_{0z}}_{=0} F_{\alpha \beta} F^{\alpha \beta}\r)_o \\
= & a_o^{-2} \l( \frac{1}{\pi} g^{\alpha \beta} F_{0 \alpha} F_{z \beta} \r)_o  
=   -a_o^{-2}\l( \frac{1}{\pi} \p_0 A_\alpha \p_z A^\alpha \r)_o \\
= &-a_o^{-2} \Bigg [ \frac{1}{\pi} \frac{\dd}{\dd\eta} \l(  \sum_{\lambda =\pm} \frac{A_I^{(\lambda)}}{r a_s} \cos \l(\omega(r -  \eta) + \varphi_I^{(\lambda)}\r) \epsilon_\alpha^{(\lambda)} \r) \\
& \cdot \underbrace{\frac{\dd r}{\dd z}}_{=-1} \frac{\dd}{\dd r} \l( \sum_{\lambda' = \pm } \frac{a_s A_I^{(\lambda')}}{r a^2(\eta)} \cos \l(\omega(r - \eta) + \varphi_I^{(\lambda')}\r) \epsilon^\alpha_{(\lambda')} \r) \Bigg ]_o
\end{align}
\begin{align}
= &  \frac{1}{\pi a_o^2}  \Bigg[ \sum_{\lambda = \pm} \sum_{\lambda'= \pm}\epsilon^{(\lambda)}_\alpha \epsilon_{(\lambda')}^\alpha \frac{A_I^{(\lambda)} A_I^{(\lambda')}}{r^2 a^2} \omega^2 \\ & \cdot \sin \l(\omega(r- \eta) + \varphi_I^{(\lambda)}\r)  \sin \l(\omega( r- \eta) + \varphi_I^{(\lambda')}\r) \Bigg]_o   + \mathcal{O}(r^{-3})\\
\simeq & \frac{1}{4 \pi r^2 a_o^4} 4 \omega^2\l[ \sum_{\lambda= \pm}\l(A_I^{(\lambda)} \sin(\omega( r-\eta) + \varphi_I^{(\lambda)})\r)^2 \r]   \\
= & \frac{1}{4 \pi r^2 a_o^4} 4 \omega_s^2 a_s^2 \l[ \sum_{\lambda= \pm}\l(A_I^{(\lambda)} \sin(a_s \omega_s ( r-\eta) + \varphi_I^{(\lambda)})\r)^2 \r]   \\
= & \frac{1}{4 \pi r^2 a_o^2(1+z)^2} 4 \omega_s^2 \l[ \sum_{\lambda= \pm}\l(A_I^{(\lambda)} \sin(a_s \omega_s( r-\eta) + \varphi_I^{(\lambda)})\r)^2 \r]  \\
= & \frac{\mathcal{L}_s}{4 \pi \l(d_L^{em}\r)^2} \,,
\end{align}
where we have defined the electromagnetic luminosity distance
\be
d_L^{em}= r a_o (1+z)  \label{eq:JordanFrameDL}
\ee
and the luminosity in the source frame
\be
 \mathcal{L}_s \equiv 4 \omega_s^2 \sum_{\lambda=\pm}\l(A_I^{(\lambda)} \sin \l( a_s \omega_s( r-\eta) + \varphi_I^{(\lambda)} \r)\r)^2 \,.
\ee
In principle, the instantaneous luminosity depends on the scale factor at the emission time.
But in practice, we average the luminosity on $\eta$ over many cycles of $\omega_s$, specifically,
\be
 \frac{1}{T}\int_0^T \mathrm{d}\eta \sin^2( a_s \omega_s(r- \eta) + \varphi_I^{(\lambda)}) = \frac{1}{2} - \frac{1}{T}\frac{\sin(2( a_s \omega_s r + \varphi_I^{(\lambda)})) + \sin(2 ( a_s \omega_s (r -T) + \varphi_I^{(\lambda)}))}{4 a_s \omega_s} \,, \label{SinIntegral}
\ee
where the second term gets suppressed after enough time and we get
\be
\frac{1}{T}\int_0^T \dd \eta \mathcal{L}_s = 2 \omega_s^2 \l(\l( A_I^{(+)}\r)^2 + \l(A_I^{(-)}\r)^2\r)\,.\label{eq:JordanFrameL}
\ee

As hinted earlier, supernovae Type Ia are no longer standard candles in the presence of a time varying $ G_{\tiny{\hbox{eff}}}(t)$. The reason is that the production mechanism of these involves a balance between the gravitational force and the electron degeneracy pressure of a white dwarf \cite{Shapiro:1983du,Amendola:1999vu}. Roughly, the emitted peak luminosity  depends on the Chandrasekhar mass $M_{Ch}$, which in turn depends on the gravitational constant through $\mathcal{L}_s \propto M_{Ch}^{\gamma}$, with $\gamma>0$, typically set to one \cite{Amendola:1999vu, Shapiro:1983du} (see however \cite{PhysRevD.97.083505} for a more sophisticated analysis showing that supernovae Type Ia light curves are still standardisable even in the presence of a time varying $ G_{\tiny{\hbox{eff}}}(t)$). The Chandrasekhar mass can be computed using hydrostatic equilibrium for a spherically symmetric stellar fluid described by a Fermi gas (see Appendix B of \cite{PhysRevD.97.083505} for example)
\be
M_{Ch}(S) =\frac{\sqrt{3 \pi}}{2} \l( \frac{\hbar c}{G_s}\r)^{3/2} \frac{1}{(\mu m_N)^2} \l( -y_s^2 \frac{\dd \Theta}{\dd y_s}\r)_{y_1} \,, \label{eq:ChandrasekharMass}
\ee
where
\be
y_s = \frac{r}{\alpha_s} \qquad \hbox{and} \qquad a_s = \sqrt{(n+1)K \lambda^{(1-n)/n}/(4 \pi G_s)}\,,
\ee
$G_s$ is the local gravitational constant at the source, $m_N$ is the nucleon mass and all quantities are defined in the Appendix of Ref.~\cite{PhysRevD.97.083505}. Here, we simply note that
\be
\alpha_s = \alpha_o \sqrt{\frac{G_o}{G_s}} \qquad \hbox{and} \qquad y_s = y_o \sqrt{\frac{G_s}{G_o}}\,,
\ee
which also implies $\dd y_s = \dd y_o \sqrt{\frac{G_s}{G_o}}$. If we are to express the Chandrasekhar mass at the source in terms of our observed quantities such as $G_o$, we get
\be
M_{Ch}(S) = \frac{\sqrt{3 \pi}}{2} \l( \frac{\hbar c}{G_o}\r)^{3/2} \l( \frac{G_o}{G_s}\r)^{3/2} \frac{1}{(\mu m_N)^2} \l( \frac{G_s}{G_o}\r)^{1/2} \l( -y_o^2 \frac{\dd \Theta}{\dd y_o}\r)_{y_1} = \l( \frac{G_o}{G_s}\r) M_{Ch}(O)\,,
\ee
where we note that the $G$ dependence differs here from Ref.~\cite{PhysRevD.97.083505}. This factor would enter through the initial amplitudes $A_I^{(\lambda)}$ in Eq. \eqref{eq:JordanFrameL} and bias the luminosity distance $d_L^{\tiny{\hbox{SNIa}}}$ inferred from a measurement of the flux 
\begin{align}
F_o^{\tiny{\hbox{SNIa}}}= \frac{\mathcal{L}_s^{\tiny{\hbox{SNIa}}}}{4 \pi \l(d_L^{em}\r)^2} \propto \frac{\l(M_{Ch}(S)\r)^\gamma}{4 \pi \l(d_L^{em}\r)^2} = \frac{\l(M_{Ch}(O)\r)^\gamma}{4 \pi  \l(\l( \frac{G_s}{G_o}\r)^{\gamma/2}d_L^{em}\r)^2} = \frac{\l(M_{Ch}(O)\r)^\gamma}{4 \pi \l(d_L^{\tiny{\hbox{SNIa}}}\r)^2} \,,
\end{align}
where we have identified
\be \label{eq:JFdLSNIa}
d_L^{\tiny{\hbox{SNIa}}} = d_L^{em}\l( \frac{G_s}{G_o}\r)^{\gamma/2} \,.
\ee 
This luminosity distance modification for supernovae Type Ia is induced by a modification in the production mechanism.

\subsection{Gravitational radiation in comparison} \label{sec:JFGWs}
Having computed the electromagnetic luminosity distance in Sec.~\ref{sec:optics}, we shall now proceed to calculate its counterpart inferred from the gravitational wave propagation and compare the two. We will also examine the emitted and received waveforms.

We first derive the equation of motion for tensor perturbations $h_{ij}(t,\mathbf{x})$ around the flat FLRW metric, $g_{ij} = a^2(t) (\delta_{ij} + h_{ij}(t,\mathbf{x}))$, travelling in the $z$-direction. The Ricci scalar contains many terms to second order in $h$, but the pure derivative terms are the only ones contributing to the wave equation in the geometric optics approximation such that the quadratic effective action reads
\be
 S^{(2)} = \frac{M_p^2}{2} \int \mathrm{d}t \mathrm{d}^3x a^3(t)G_4(\phi(t)) \l( \frac{1}{4} \dot{h}_{ij} \dot{h}_{kl} - \frac{\delta^{mn}}{4 a^2(t)} \p_m h_{ij} \p_n h_{kl}  \r)\delta^{lj} \delta^{ki} \,,
\ee
where $t$ denotes cosmological time.
Variation of this action with respect to $h_{ij}$ and its first derivative leads to the Euler-Lagrange equations
\be
 \frac{\p (\sqrt{-g} \mathcal{L}^{(2)})}{\p h_{ij}} - \p_\mu \frac{\p (\sqrt{-g} \mathcal{L}^{(2)})}{\p(\p_\mu h_{ij})} =0
\ee
and hence the gravitational wave equation
\be \label{eq:GWJordan}
 \p_\eta^2 h_{ij} + \mathcal{H} (2 + \aM) \p_\eta h_{ij} - \vec{\nabla}^2 h_{ij} =0 \,,
\ee
where $h_{11} = h_{22}=  h_+$, $h_{12} = h_{21} = h_\times$, and~\cite{Bellini:2014fua}
\be \label{alphaM}
\aM \equiv \mathcal{H}^{-1} \p_\eta \log M^2
\ee
denotes the temporal rate of change of the squared effective Planck mass $M^2 = M_p^2 G_4(\phi)$.
Since the two polarisations decouple and satisfy the same wave equation, we will simply refer to both as $h_A$.
To solve  Eq.~\eqref{eq:GWJordan}, we assume that the wave propagates radially, $h_A = \frac{1}{r} f_A(r,\eta)$, such that
$\p_\eta^2 f_A + \mathcal{H} (2 + \aM) \p_\eta f_A - \p_r^2 h_A =0$,
or in Fourier space,
\be
\p_\eta^2 f_A + \mathcal{H} (2 + \aM) \p_\eta f_A + k^2 h_A =0 \,.
\ee
To absorb the friction term, we perform the substitution \cite{Belgacem:2018lbp}
\be
 f_A(\eta,k) = \frac{1}{a_*(\eta)} \chi_A(\eta,k) \,,
\ee
where we define an effective scale factor $a_*$ via
\be \label{eq:atilde}
 \frac{a_*'}{a_*} \equiv \mathcal{H}\l(1+ \frac{\aM}{2} \r) \,.
\ee
This yields $\chi_A''(\eta,k) + \l(k^2 -\frac{a_*''}{a_*} \r )\chi_A(\eta,k)=0$\,.
Neglecting $\frac{a_*''}{a_*} \ll k^2$ for modes that are well inside the horizon, one finds that $\chi$ simply satisfies the standard wave equation
\be
 \chi_A''(\eta,k) + k^2 \chi_A(\eta,k)=0 \,.
\ee
Now, the difference to a gravitational wave in GR lies in the fact that $h_A(\eta,k)$ decreases with $1/a_*$ for $\aM$ instead of $1/a$.
Therefore, the GW amplitude observed today after propagating from the source at redshift $z$ to the observer has decreased by a factor
$a_*(z)/a_*(0) $ instead of $a(z)/a(0)$. We shall assume that spatial variations of the gravitational coupling at the source, $G_{\tiny{\hbox{eff}}}$, are suppressed~\cite{Lombriser:2015sxa,Belgacem:2019pkk}, which may be expected for a viable gravitational modification in a high-density regime and is certainly realised when the source is screened.
Note, however, that we allow for $G_{\tiny{\hbox{eff}}}$ to vary in time.
When transforming the gravitational wave amplitudes from the frame of the source to that of the observer, we obtain
\begin{align}
 h_+ (\tau_s) = & \frac{a_*(\tau_s)}{a_*(\tau_o)}  \frac{4 (G_{\tiny{\hbox{eff}}}(\tau_s) M_c)^{5/3}}{a(\tau_s)r} \l( \pi f_s\r)^{2/3} \frac{1 + \cos^2 \imath}{2} \cos \l( \Phi_s(\tau_s)\r) \\
 = & \frac{a_*(\tau_s)}{a(\tau_s)} \frac{4 (G_{\tiny{\hbox{eff}}}(\tau_s) \mathcal{M}_c)^{5/3}}{d_L^{em}} \l( \pi f_o\r)^{2/3} \frac{1 + \cos^2 \imath}{2} \cos \l( \Phi_s(\tau_s)\r),
\end{align}
where we set $a(\tau_o)=1= a_*(\tau_o)$ without loss of generality since only the ratios of $a$ and $a_*$ appear. We also switch the frequency at the source to that in the frame of the observer, $f_s = (1+z)f_o$ and change to the redshifted chirp mass $\mathcal{M}_c(z) = (1+z) M_c$~\cite{Maggiore:1900zz}, where the luminosity distance $d_L^{em}$ is given by Eq.~\eqref{eq:JordanFrameDL}.

Let us briefly inspect the effects on the waveform by an effective Planck mass varying on cosmological time scales. To lowest order, the motion of the binary system is effectively described by GR with a change in the Planck mass.
For an event at a cosmological distance, the effective Planck mass at the emission time will in general be different from that at the observing time due to the the slowly-varying $M^2$, but during the inspiraling phase, taking up to a few minutes for the observed binary neutron star, $M^2$ remains effectively constant, thus, $G_{\tiny{\hbox{eff}}}(z)=cst$. Therefore, the only difference with respect to GR is the friction term in Eq.~\eqref{eq:GWJordan} and the related effective Newton constant.
At lowest post-Newtonian (PN) order in the TT gauge, the waveforms satisfy 
\begin{align}
 h_+(\tau_o) = & h_c(\tau_o) \frac{1 +\cos^2 (\imath)}{2} \cos \l( \Phi (\tau_o)\r) \,, \\
 h_\times(\tau_o) = & h_c(\tau_o) \cos( \imath) \sin(\Phi (\tau_o))
\end{align}
with
\begin{align}
 \Phi(\tau_o) = & -2 \l( \frac{5}{c^3}  G_{\tiny{\hbox{eff}}}(0)  \frac{G_{\tiny{\hbox{eff}}}(z)}{G_{\tiny{\hbox{eff}}}(0)}\mathcal{M}_c(z)\r)^{-5/8} \tau_o^{5/8} + \Phi_0 \\
 = &-2  \l( \frac{5}{c^3} G_{\tiny{\hbox{eff}}}(0)  \l( \frac{C(0)}{C(z)}\r)^2 \mathcal{M}_c(z)\r)^{-5/8} \tau_o^{5/8} + \Phi_0 \label{eq:JordanFramePhase}
\end{align}
and
\begin{align} \label{eq:JordanFrameAmplitude}
 h_c(\tau_o) = & \frac{4}{d_L^{em} \frac{a(z)}{a_*(z)}} \l( G_{\tiny{\hbox{eff}}}(O) \l(\frac{C(0)}{C(z)}\r)^2 \mathcal{M}_c(z)\r)^{5/3} \l( \pi f_{gw}^{(o)}(\tau_o) \r)^{2/3} \,,
\end{align}
where we have made use of the conformal coupling $C=\sqrt{G_4}$, which will become useful in Sec.~\ref{sec:EframeGW} when comparing results to those in the Einstein frame.
The chirp mass is measured by comparison of the measured frequency to its time variation,
\be
\frac{\dd f_{gw}^{(s)}}{\dd \tau_s} = \frac{96}{5} \pi^{8/3} \l( G_{\tiny{\hbox{eff}}}(z) M_c \r)^{5/3} \l[ f_{gw}^{(s)}\r]^{11/3} \,.
\ee
In the frame of the observer with $\dd \tau_o = (1+z) \dd \tau_s$ and $f_s = f_o (1+z)$~\cite{Maggiore:1900zz}, we find
\begin{align}
\frac{\dd f_{gw}^{(o)}}{\dd \tau_o} =  &\frac{96}{5} \pi^{8/3} \l( G_{\tiny{\hbox{eff}}}(0) \l(\frac{C(0)}{C(z)}\r)^2 M_c \r)^{5/3} (1+z)^{5/3} \l[ f_{gw}^{(o)}\r]^{11/3} \\
 & = \frac{96}{5} \pi^{8/3} \l( G_{\tiny{\hbox{eff}}}(0) \l(\frac{C(0)}{C(z)}\r)^2 \mathcal{M}_c(z) \r)^{5/3} \l[ f_{gw}^{(o)}\r]^{11/3} \,.
\end{align}
From the measurement $f_{gw}^{(o)}(\tau_o)$ and its time variation, we therefore infer the effective redshifted chirp mass
\be \label{eq:EffectiveChirpMass}
\mathcal{M}_{\tiny{\hbox{c,eff}}}(z)\equiv \l( \frac{C(0)}{C(z)}\r)^2 \mathcal{M}_c(z) =  \l( \frac{C(0)}{C(z)}\r)^2 (1+z) M_c \,,
\ee
which appears in Eq.~\eqref{eq:JordanFramePhase} and Eq.~\eqref{eq:JordanFrameAmplitude}. From Eq.~\eqref{eq:EffectiveChirpMass} it is clear that the variation of the Planck mass from the redshift of the source to the observer, or of the conformal factor, is degenerate with the chirp mass. Hence, the only discriminable effect for the cosmological modification of gravity on the inspiral phase is that on the distance. Note however that an inspiral-merger-ringdown analysis may break the degeneracy between the effective chirp mass and the actual chirp mass~\cite{TheLIGOScientific:2016wfe}. One may define a gravitational wave luminosity distance as
\be
 d_L^{gw}(z) \equiv \frac{a(z)}{a_*(z)} d_L^{em}(z) \,.
\ee
Standard Sirens thus measure $d_L^{gw}(z)$ rather than $d_L^{em}(z)$~\cite{Saltas:2014dha,Lombriser:2015sxa}.
Solving Eq.~\eqref{eq:atilde} for $a_*$, we find~\cite{Amendola:2017ovw,Belgacem:2017ihm,Linder:2018jil,Belgacem:2019pkk}
\be \label{eq:DLgwtoDLem}
 R(z)\equiv \frac{d_L^{gw}(z)}{d_L^{em}(z)}= \exp \l (\frac{1}{2} \int_0^z \frac{\aM(z)}{(1+z)}\mathrm{d}z \r ) = \frac{M(0)}{M(z)} = \frac{C(0)}{C(z)} = \sqrt{\frac{G_{\tiny{\hbox{eff}}}(z)}{ G_{\tiny{\hbox{eff}}}(0)}} \,,
\ee
where we have also written the ratio in terms of the effective gravitational coupling $G_{\tiny{\hbox{eff}}}(z) \equiv M^{-2}(z)$. In GR, we have $\aM=0$ and this equation reduces to $d_L^{gw}(z) = d_L^{em}(z)$. In contrast, for general $\aM \neq0$, $R(z)\neq1$, which allows for a test of the nonminimal coupling of the scalar field to the Ricci scalar.
However, from our derivation from Eq.~\eqref{eq:GWJordan} it is ambiguous whether the Planck masses in Eq.~\eqref{eq:DLgwtoDLem} should be interpreted as those of the cosmological background at the time of emission and observation or the ones local to the source and observer (also see Ref.~\cite{Amendola:2017ovw}).
While this may potentially not make a difference for Vainshtein screening (Sec.~\ref{sec:horndeski}), it is crucial for any observationally compatible scenario, where the time evolution of the Planck mass must be locally screened.

Importantly, if the relevant coupling is indeed the local one (as we will find), the local evolution of the Planck mass should also be considered for the electromagnetic sources.
As discussed in Sec.~\ref{sec:optics}, the emitted luminosity of supernovae Type Ia would accordingly be modified, and as a result the ratio $R(z)$ would instead amount to
\be
R(z) = \frac{d_L^{gw}}{d_L^{\tiny{\hbox{SNIa}}}}=  \l(\frac{C_o}{C_s}\r)^{ 1 -\gamma } \,, \label{eq:DLgwtoDLSNIa}
\ee
where the power of $C_o/C_s$ depends on $\gamma>0$, which parametrises the dependence of the peak luminosity on the Chandrasekhar mass.
We highlight that for a linear dependence of $\mathcal{L}_s$ on $M_{Ch}$ (i.e., $\gamma=1$), this ratio of the luminosity distances for modified gravity reduces to one, as in GR, even in the absence of a screening mechanism for the time variation of $G_{\tiny{\hbox{eff}}}$.
We will further address the question surrounding a Planck mass evolution locally or in the background in Sec.~\ref{sec:screeningmechanism}.
But before doing so, we shall first confirm our results with an independent calculation in the Einstein frame.

\section{Light and gravitational wave propagation in the Einstein frame}\label{sec:EinsteinFrame}

In Sec.~\ref{sec:JordanFrame}, we have derived the photon and gravitational wave equations in the Jordan frame of Horndeski scalar-tensor theories that satisfy $\cT=1$ and we have classified the observational difference between the luminosity distances inferred from photons and gravitational waves.
We now wish to perform these calculations in the Einstein frame.
The motivation for such a calculation is threefold: we want to check the consistency of the theoretical prediction for the modified gravity signature in Standard Sirens, the formulation in Einstein frame allows for an intuitive implementation of screening effects (Sec.~\ref{sec:screeningmechanism}), and the computations in Einstein frame enable a generalisation of the effects to non-universal couplings (Sec.~\ref{sec:darksectorinteractions}).
We will first briefly review the relevant computational aspects of changing the frame in Sec.~\ref{sec:framechange}, deriving the photon and tensor equations of motion in Sec.~\ref{sec:EFEoMs}.
In order to compute the flux in Sec.~\ref{sec:EFFlux}, we will first show that photons follow null geodesics in the Einstein frame (Sec.~\ref{sec:EFgeodesics}) and inspect the effect of the frame transformation on spinor fields in curved spacetime in Sec.~\ref{sec:spinors}.
Importantly, in Sec.~\ref{sec:EFFlux} we will also put special care in the correct notion of redshift.
With these preparations, we show the frame invariance of the electromagnetic luminosity, the distance inferred from it, the Chandrasekhar mass and the supernovae Type Ia luminosity distance.
Finally, in Sec.~\ref{sec:EframeGW} we also show the frame invariance of the gravitational wave luminosity distance and therefore of its ratio to the electromagnetic counterpart providing the observable signature of modified gravity.

\subsection{Change of frame} \label{sec:framechange}

The Einstein-frame action is obtained from the Jordan-frame action \eqref{Horndeski_Action} by transforming the minimally coupled metric $g_{\mu\nu}$ to a new, nonminimally coupled metric $\tilde{g}_{\mu\nu}$ that recovers the standard Einstein-Hilbert action in terms of $\tilde{g}_{\mu\nu}$.
We shall only consider conformal transformations here,
\be \label{eq:metrictransformation}
 g_{\mu\nu}(x)\to \tilde{g}_{\mu\nu}(x) = C^2(x)g_{\mu\nu}(x) \,,
\ee
as universal disformal couplings give rise to observationally incompatible gravitational wave speeds $\cT\neq1$ (Sec.~\ref{sec:horndeski}), but we will consider the presence of disformal couplings when discussing dark sector interactions in Sec.~\ref{sec:darksectorinteractions}. In general, tildes will denote quantities in the Einstein frame, for which indices are raised and lowered with $\t{g}$.
Eq.~\eqref{eq:metrictransformation} implies
\be
 \tilde{g}^{\mu\nu} = C^{-2}g^{\mu\nu} \,, \quad \sqrt{-\tilde{g}} = C^{4}\sqrt{-g} \,.
\ee
Conformal transformations are different than regular coordinate transformations as the spacetime invariant $\dd s^2$ is also affected by the transformation.
The Jordan and Einstein frame spacetime invariants are related by
\be
 \dd \t{s}^2 = \t{g}_{\mu\nu} \dd x^\mu \dd x^\nu = C^2 g_{\mu\nu} \dd x^\mu \dd x^\nu = C^2 \dd s^2 \,.
\ee
For an FLRW universe, one finds
\be\label{eq:spacetimeinvariant}
 \dd \t{s}^2 = C^2(x) g_{\mu\nu} \dd x^\mu \dd x^\nu = -C^2(x) \dd t^2 + C^2 a^2 \dd \mathbf{x}^2 = - \dd \t{t}^2 + \t{a}^2(\t{t})\dd \mathbf{x}^2 \,,
\ee 
where we have defined a new spacetime dependent time coordinate $\dd\t{t} = C(t,\mathbf{x})\dd t$, an Einstein frame scale factor $\t{a}(t, \mathbf{x})= C(t,\mathbf{x})a(t)$, and left the space coordinates unchanged. Conformal time is invariant under conformal transformation as long as we absorb the conformal factor in the scale factor.
It is a bit of a lengthy calculation, but one finds (see, e.g., Ch.~3.2 in Ref.~\cite{Fujii:2003pa})
\begin{align}
 R = C^2(x) \l( \t{R} + \dots \r) \,.
\end{align}
The gravitational part of action~\eqref{eq:horndeski} (after applying $\cT=1$~\cite{McManus:2016kxu}) that also contains the quadratic terms in $h_A$ transforms as
\begin{align}
 S &= \frac{M_p^2}{2}\int \dd^4 x \sqrt{-g} \left ( G_4(\phi) R + \dots \right ) = \frac{M_p^2}{2}\int  \dd^4 x \sqrt{-\tilde{g}} C^{-4}  \left ( G_4(\phi) C^2 \tilde{R} + \dots \right ) \\ &=  \frac{M_p^2}{2} \int \dd^4 x \sqrt{-\tilde{g}}  \left (\tilde{R} + \dots \right ) \,,
\end{align}
where we have set the conformal factor as $C^2(x) = G_4(\phi)$, justifying the choice in Eq.~\eqref{eq:DLgwtoDLem}.
After this transformation, the nonminimal coupling of $\phi$ to the Ricci scalar in the gravitational part of the action has disappeared.
We are left with the Einstein-Hilbert action for $\t{g}$, which leads to the Einstein field equations of GR.
But we now also have a nonminimal coupling of $\phi$ in the matter sector, which shall be our focus in the following.

We emphasise that the two frames are physically equivalent but the interpretations in each frame might vary from one another.
For example, the Planck mass run rate $\aM$ becomes a rate of change of the conformal factor,
\begin{align}
 \aM &= \mathcal{H}^{-1} \p_\eta \ln M^2 = \mathcal{H}^{-1} \p_\eta \ln 2 G_4 = 2 \mathcal{H}^{-1} \frac{C'}{C} \,.
\end{align}

\subsection{Photon and tensor equations of motion} \label{sec:EFEoMs}

To inspect the photon propagation in the Einstein frame, let us first recast the electromagnetic action~\eqref{eq:EMaction} accordingly.
This gives
\begin{align} \label{eq:EMActionEinstein}
 S_{EM} &= -\frac{1}{4\pi}\int \dd^4 x \sqrt{-g} F_{\mu\nu}F^{\mu\nu} = -\frac{1}{4 \pi}\int \dd^4 x \sqrt{-g} g^{\mu\alpha}g^{\nu\beta} F_{\mu\nu}F_{\alpha \beta} \\ &= -\frac{1}{4 \pi}\int \dd^4 x \sqrt{-\tilde{g}}C^{-4} C^2 \tilde{g}^{\mu \alpha}C^2 \tilde{g}^{\nu\beta} F_{\mu\nu}F_{\alpha\beta}= -\frac{1}{4 \pi } \int \dd^4 x \sqrt{-\tilde{g}} \tilde{g}^{\mu \alpha} \tilde{g}^{\nu\beta} F_{\mu\nu}F_{\alpha\beta} \\ &= -\frac{1}{4\pi}\int \dd^4 x \sqrt{-\tilde{g}} \tilde{g}^{\mu \alpha} \tilde{g}^{\nu\beta} F_{\mu\nu}F_{\alpha\beta} \,,
\end{align}
where due to the symmetric Christoffel symbols, $F_{\mu\nu} = \nabla_\mu A_\nu - \nabla_\nu A_\mu = \p_\mu A_\nu - \p_\nu A_\mu$ is independent of the metric.
This independence motivates the lowering of the indices of the field strength before performing the transformation, which hence must be done with caution.
Since the transformed action is exactly that of Eq.~\eqref{eq:EMaction} with $g_{\mu\nu}$ replaced by $\tilde{g}_{\mu\nu}$, in the Lorenz gauge ($\t{\nabla}_\mu A^\mu=0$), we find the same equation of motion as Eq. \eqref{curvedPhotonEoM} with $\nabla$ replaced by $\tilde{\nabla}$ and $R$ replaced by $\t{R}$:
\begin{align}
 \tilde{\nabla}_{\mu}\tilde{\nabla}^\mu A^\nu - \te{\t{R}}{^\nu _\mu} A^\mu &= 0 \,, \label{eq:photonsEoMEF1} \\
 \tilde{\nabla}_\mu\tilde{\nabla}^\mu A_\sigma - \te{\t{R}}{_\sigma^\rho} A_\rho  &=0 \label{eq:photonsEoMEF2} \,.
\end{align}
For the gravitational radiation, given the recovery of the standard Einstein-Hilbert term in the Einstein frame, the wave equation is that of GR with 
\be
 \t{\nabla}^\rho \t{\nabla}_\rho \t{h}_{\mu\nu} + 2 \te{\t{R}}{_\mu _\rho _\nu _\sigma} \t{h}^{\rho \sigma} = 0 \,. \label{eq:curvedGWEoM}
\ee
For the FLRW metric,
\be
 \t{g}_{ij}(\eta,\mathbf{x}) = \t{a}^2(\eta,\mathbf{x}) (\delta_{ij} + \t{h}_{ij}(\eta,\mathbf{x}))= C^2(\eta,\mathbf{x}) a^2(\eta) (\delta_{ij} + h_{ij}(\eta,\mathbf{x})) = C^2(\eta,\mathbf{x}) g_{ij}(\eta,\mathbf{x})
\ee
and hence the conformal factor is absorbed in the scale factor such that $\t{h}_{ij}= h_{ij}$. Explicitly writing the covariant derivatives in Eqs.~\eqref{eq:photonsEoMEF1}--\eqref{eq:curvedGWEoM} in terms of the FLRW Christoffel symbols and evaluating these on the FLRW background, we get in the Lorenz gauge and in the TT gauge for the electromagnetic and gravitational wave equations,
\begin{align}
&\p_\eta^2 A_\sigma - \vec{\nabla}^2 A_\sigma =0 \,, \label{eq:A1} \\
&\p_\eta^2 A^\nu + 4 \tilde{\mathcal{H}} \p_\eta A^\nu - \l( \vec{\nabla}^2 - 2 \frac{\t{a}'' \t{a} + (\t{a}')^2}{\t{a}^2}\r) A^\nu=0 \,, \label{eq:A2} \\ 
&\partial_\eta^2 \t{h}_{ij} + 2 \tilde{\mathcal{H}} \partial_\eta \t{h}_{ij} - \vec{\nabla}^2 \t{h}_{ij} =0 \,, \label{eq:GWEinsteinFrameEoM}
\end{align}
respectively.
It is interesting that for an FLRW background, the curvature terms in Eq.~\eqref{eq:curvedGWEoM} cancel exactly to yield Eq.~\eqref{eq:GWEinsteinFrameEoM}.
Note that one can alternatively also use the quadratic effective action to obtain Eq.~\eqref{eq:GWEinsteinFrameEoM}.

The $A^\nu$ field carries twice the damping of the gravitational wave $\t{h}_{ij}$.
This is just as expected for GR, where the factor of two arises due to measuring the flux of light $F\propto (d_L^{em})^{-2}$ whereas $d_L^{gw}$ is probed by the amplitude of $\t{h}_{ij} \propto (d_L^{gw})^{-1}$ (Sec.~\ref{sec:JordanFrame}).
Given propagation equations as in GR, one would na\"ively expect that the luminosity distance inferred for light and for gravitational waves  must therefore always be the same in the Einstein frame, contrary to the prediction in the Jordan frame (Sec.~\ref{sec:JFGWs}), where the wave equations
showed explicitly that the damping effects differ.
Importantly, however, the equations of motion of the electromagnetic potential are not directly observable, and we will need to compute the effect on the observable quantities (Secs.~\ref{sec:EFFlux} and \ref{sec:EframeGW}).
%

\subsection{Geodesics} \label{sec:EFgeodesics}

The matter fields follow geodesics of the nonminimally coupled metric $g$ (i.e., $k^\mu \nabla_\mu k^\nu =0$).
We now want to find the analogue expression in terms of the metric $\t{g}$, for which we obtain
\begin{align}
 k^\mu \t{\nabla}_\mu k^\nu = & k^\mu \l( \p_\mu k^\nu + \t{\Gamma}^\nu_{\mu\lambda} k^\lambda \r) \\
 = & \underbrace{k^\nu \nabla_\mu k^\nu}_{=0} + k^\mu \l( \delta_\mu^\nu \frac{\p_\lambda C}{C} + \delta^\nu_\lambda \frac{\p_\mu C}{ C} - g_{\mu\lambda} \frac{g^{\nu \sigma}\p_\sigma C}{C} \r) k^\lambda \\
 = & 2 k^\mu \frac{\p_\mu C}{ C} k^\nu - k^\mu k_\mu C \t{g}^{\nu\sigma}\p_\sigma C \,. \label{eq:EinsteinGeodesics}
\end{align}

For massless particles such as photons and gravitons in the geometric optics approximation, we are allowed to absorb the apparent fifth force in the affine parameter, $\mathrm{d} \t{\lambda} = C^2(x) \mathrm{d}\lambda$~\cite{Wald:1984rg}, which does not bear a physical meaning.
Then Eq.~\eqref{eq:EinsteinGeodesics} becomes,
\be
 \t{k}^\mu \t{\nabla}_\mu \t{k}^\nu = - \t{k}^\mu \t{k}_\mu \frac{\p^\nu C}{C}
\ee
with $\t{k}^\mu = \dd x^\mu /\dd \t{\lambda}$.
Because of the invariance of the electromagnetic action under a conformal transformation (Sec.~\ref{sec:EFEoMs}), photons still satisfy the null geodesic equation $\t{k}^\mu \t{k}_\mu =0$ and therefore photons still follow null geodesics in Einstein frame,
\be
\t{k}^\mu \t{\nabla}_\mu \t{k}^\nu =0 \,.
\ee
Of course, this equation is invariant under a change of affine parameter of the type $\lambda_* = a \t{\lambda} + b$ with $a \neq 0 $, $a,b \in \mathbb{R}$, as $\mathrm{d}\lambda_* = a \mathrm{d}\t{\lambda}$.
The relation for the four-vector of photons between the Einstein and Jordan frames, up to a constant factor of proportionality, is
\be
 k^\mu = \frac{\dd x^\mu}{\dd \lambda} = \frac{\dd \t{\lambda}}{ \dd \lambda}  \frac{\dd x^\mu}{\dd \t{\lambda}} \propto  C^2 \t{k}^\mu \,,
\ee
and so
\be
 \t{k}_\mu = \t{g}_{\mu\nu} \t{k}^\nu \propto C^2 g_{\mu\nu}  C^{-2} k^\nu = k_\mu \,.
\ee
The geodesic equation implies
\be
 \frac{\dd \t{k}^\mu}{\dd \t{\lambda}} = \frac{1}{2} (\t{g}_{\alpha \kappa, \mu}) \t{k}^\alpha \t{k}^\kappa \,.
\ee
For a homogeneous space such as a FLRW universe, $\t{g}_{\alpha \kappa, i} =0$, we have $\t{k}_\mu = cst$, which is enforced by $\t{k}_\mu \t{k}^\mu=0$.
The same also holds true in the Jordan frame such that $\t{\omega} \propto \omega = \hbox{cst}$.
Note that it is not a problem that the affine parameter $\t{\lambda}$ is related to $\lambda$ via the conformal factor $C^2$, which depends on the spacetime point, as long as the sign of $C^2$ is preserved, which is naturally expected for gravity. 
The only thing that matters is that $\t{\lambda}$ parametrises the curve and that it is monotonic from the source to the observer.
The freedom that we have to absorb the conformal factor is not possible for matter, where the affine parameter is the proper time, which reserves a special transformation rule under conformal transformations.

For massive particles, we return to Eq.~\eqref{eq:EinsteinGeodesics}, hence,
\begin{align}
 k^\mu \t{\nabla}_\mu k^\nu = & 2 k^\mu \frac{\p_\mu C}{ C} k^\nu - k^\mu k_\mu C\t{g}^{\nu\sigma}\p_\sigma C \,. \label{eq:EFmassivegeodesic}
\end{align}
Since $k^\mu = m u^\mu$ and every term contains two factors of $k^\mu$, we simplify the expression by extracting $m^2$ and replace $k^\mu$ with $u^\mu$.
We cannot use here the freedom of absorbing the fifth force in the affine parameter since proper time is observable.
We have
\be
 u^\mu = \frac{\dd x^\mu}{\dd \tau} = \frac{\dd \t{\tau}}{\dd \tau} \frac{\dd x^\mu}{\dd \t{\tau}} = C \t{u}^\mu
\ee
and
\be
 u_\nu = C^{-1} \t{u}_\nu \,,
\ee
which when used in Eq.~\eqref{eq:EFmassivegeodesic} gives
\be
 C \t{u}^\mu \t{\nabla}_\mu (C \t{u}^\nu) = 2 \frac{\p_\mu C}{C} C^2 \t{u}^\mu \t{u}^\nu - C \t{u}^\mu C^{-1} \t{u}_\mu C \t{g}^{\nu\lambda} \p_\lambda C \,,
\ee
simplifying to
\be
 \t{u}^\mu \t{\nabla}_\mu \t{u}^\nu = \l( \frac{\p_\mu C}{C}  \t{u}^\nu \t{u}^\mu -   \frac{\p_\lambda C}{C}\t{g}^{\nu \lambda}  \t{u}_\mu  \t{u}^\mu \r) \,.
\ee
The term on the right-hand side is typically referred to as the fifth force, i.e., a massive particle does not follow the geodesic of the metric of that frame.
Note that if we assume $\t{u}^i=0$ from the start, we have
\be
 \t{u}^0 \p_0 \t{u}^0 + \t{\Gamma}^0_{ij}\underbrace{\t{u}^i \t{u}^j}_{=0} = \l( \frac{\p_0 C}{C}  \t{u}^0 \t{u}^0 -   \frac{\p_0 C}{C} \underbrace{\t{g}^{00}}_{=-1}  \underbrace{\t{u}_0}_{=-\t{u}^0}  \t{u}^0 \r) =0 \,,
\ee
which leads to $\t{u}^0 = \frac{\dd\t{\tau}}{\dd\t{\tau}}= 1$, provided $\t{u}^\mu \t{u}_\mu =-1$.
In conformal time this reads $\t{u}^0 = \frac{\dd \eta}{\dd \t{\tau}} = \t{a}^{-1}$.
Therefore, the fifth force does not act on objects already at rest in the Jordan frame.
They remain at rest in the Einstein frame.

\subsection{Spinor fields in curved spacetime} \label{sec:spinors}

In order to meaningfully interpret observations in the Einstein frame and since photons interact with matter, in particular being emitted and received by matter, it is important to understand the transformation rules for the matter fields at the fundamental level described by spinors.
Hence, we will now inspect how the spinors transform under a conformal transformation.
These propagate here in curved spacetime and we will have to make a few distinctions with respect to spinors defined on flat Minkowski space. For this purpose, we may relate the metric $g_{\mu\nu}$ to the Minkowski metric $\eta_{ab} = \hbox{diag}(-1,1,1,1)$ adopting the vierbein $e_\mu^a$ formalism with
\be
 g_{\mu\nu} = e_\mu^a e_\nu^b \eta_{ab} \,,
\ee
where Latin indices are raised and lowered using $\eta_{ab}$ whereas Greek indices are raised and lowered using $g_{\mu\nu}$.
For example, $e^{\mu a} = g^{\mu\nu} e_\nu^a$ or $e_{\nu a} = \eta_{ab} e_\nu^b$.
The spin connection is defined as
\be
 \omega_\mu^{ab} = e_\nu^a \l(\p_\mu e^{\nu b} + \Gamma^\nu_{\sigma \mu} e^{\sigma b}\r) \,.
\ee
The action for a spinor field interacting with a vector field is given by (e.g., Ref.~\cite{Domenech:2016yxd})
\be
 S= \int \mathrm{d}^4 x \sqrt{-g} \l ( - i \overbar{\psi} \gamma^\mu D_\mu \psi - m \overbar{\psi} \psi - \frac{1}{4} F_{\mu\nu} F^{\mu\nu}\r) \,,
\ee
where $D_\mu = \p_\mu + i e A_\mu - \frac{1}{8} \omega_{ab \mu} [\gamma^a, \gamma^b]$, $\gamma^a$ are the Dirac matrices satisfying the flat space Clifford algebra $\{ \gamma^a, \gamma^b\} = -2 \eta^{ab}$ and $\gamma^\mu = e^\mu_a \gamma^a$.
In Ref.~\cite{Domenech:2016yxd}, it was shown that under the conformal transformation $\t{g}_{\mu\nu} = C^2 g_{\mu \nu}$, provided we can neglect derivative interactions of the scalar field with the spinor fields, the action becomes
\be
 S= \int \mathrm{d}^4x \sqrt{-\t{g}} \l( - i \overbar{\t{\psi}} \t{\gamma}^\mu D_\mu \t{\psi} - \t{m}\overbar{\t{\psi}} \t{\psi} - \frac{1}{4} F_{\mu\nu} F^{\mu\nu} \r) \,,
\ee
where $\t{\gamma}^\mu = C^{-1} \gamma^\mu$, $\t{\psi} = C^{-3/2} \psi$, $\t{m}= C^{-1}m$.
$A_\mu$ and the electron charge $e$ are conformally invariant. Since the Bohr radius $\t{a}_0$ is inversely proportional to the electron mass, this implies
\be
 \t{a}_0 = \frac{\hbar}{\t{m}_e c \alpha} \,.
\ee
Hence, the Bohr radius in Einstein frame becomes spacetime dependent $\t{a}_0(t, \mathbf{x}) = C(t,\mathbf{x}) a_0$.
Consequently, the transition lines from an $n_i$ state to an $n_f$ state of a Hydrogen atom also become spacetime dependent,
\be
 \t{E}_{n_i n_f}(t,\mathbf{x}) = \t{R}_\infty(t,\mathbf{x}) \l( \frac{1}{n_i^2} - \frac{1}{n_{f}^2} \r) = \frac{\alpha^2 \t{m}_e(t,\mathbf{x})}{4 \pi \hbar}\l( \frac{1}{n_i^2} - \frac{1}{n_{f}^2} \r) =C^{-1}(t,\mathbf{x}) E_{n_i n_f} \,.
\ee
We choose the normalisation such that the emission lines measured by the observer are frame independent,
\be
\t{E}_{n_i n_f}(t, \mathbf{x}) = \frac{C(t_o,\mathbf{x_o})}{C(t,\mathbf{x})} E_{n_i n_f} \,,
\ee
which constitutes an important step towards understanding redshift in Einstein frame (Sec.~\ref{sec:EFFlux}).

\subsection{Flux and redshift} \label{sec:EFFlux}

We now proceed to calculating the luminosity distance in the Einstein frame $\t{d}_L^{em}$.
Since the luminosity distance is an observable quantity it should not be affected by the frame transformation.
As we will see, it will require nontrivial cancellations arising from the damping term in the electromagnetic wave equations with the four-velocity and geometry around the observer to confirm this invariance and consistently recover the observed luminosity distance $d_L^{em}$ as predicted in the Jordan frame. The four-velocities of the observer and of the source in the Einstein frame are obtained from
\be
\t{u}^\mu \equiv \frac{\dd x^\mu}{\dd \t{\tau}}= C^{-1} u^\mu \,.
\ee
Since the electromagnetic action~\eqref{eq:EMActionEinstein} is conformally invariant, the same equations of motion for electromagnetism can be derived and one can easily show that photons follow null geodesics (Sec.~\ref{sec:EFEoMs}).
The geodesic equation implies that
\be
 \frac{\dd \t{k}_\mu}{\dd \t{\lambda}} = \frac{1}{2} \l(\t{g}_{\alpha \kappa, \mu}\r) \t{k}^\alpha \t{k}^\kappa \,.
\ee
Consider now a uniform scalar field $\phi(t)$ for each time slice of the universe (see Sec.~\ref{sec:screeningmechanism} for spatial dependencies) such that $C(x)=C(z(t))$. For an Einstein frame metric $\t{g}$ that is then independent of space coordinates such as the flat FLRW metric, we must have $\t{k}_i = \hbox{cst}$, along geodesics, which implies $\t{k}_\nu(O) = (-\t{\omega},0,0,\t{\omega})$, and $\t{k}_\nu(S) = (-\t{\omega},0,0,\t{\omega})$, with $\t{\omega} = \hbox{cst}$.
We can then choose the affine parameter such that $\dd \t{\lambda} = \frac{C^2}{C_o} \dd \lambda$.
The $C^2$ absorbs the apparent fifth force on photons and the $C_o$ is allowed by an affine redefinition of the affine parameter (see Sec.~\ref{sec:EFgeodesics}). This factor is chosen such that $\t{\omega}_o = \omega_o$ in Eq.~\eqref{eq:unscreened_frequencies_wO}.
We arrive at $\t{k}_\mu = C_o k_\mu$, which preserves the null geodesic equation for photons and we find $\t{k}^\mu = C_o C^{-2}k^\mu$ and $\t{\omega} =C_o \omega= \hbox{cst}$.
Thus,
\begin{align}
 \t{\omega}_s &=  \frac{C_o}{C_s} \omega_s \,, \label{eq:unscreened_frequencies} \\
 \t{\omega}_o &= \omega_o \label{eq:unscreened_frequencies_wO}
\end{align}
such that
\be
 \frac{\t{\omega}_s}{\t{\omega}_o} = \frac{C_o}{C_s}(1+z) \label{eq:nonredshift} \,.
\ee
From Eq.~\eqref{eq:nonredshift} it would na\"ively seem that the redshift of photons is affected by the conformal tranformation. Recall, however, that in the Einstein frame, the Bohr radius is evolving with time, and hence that we require a more
careful definition of what we refer to as the redshift.

More accurately, the redshift $z$ of a transition line from an $n_i$ to an $n_f$ level of a given atom is the deviation from unity in the ratio between the frequency emitted that we measure in the lab $\t{E}_{n_i n_f}(O)$ from this transition and the one that we observe from a distant galaxy $\t{\omega}_o$ such that
\be \label{eq:redshift}
(1+\t{z}) \equiv \frac{\t{E}_{n_i n_f}(O)}{\t{\omega}_o} = \frac{\frac{C_s}{C_o}\t{E}_{n_i n_f}(S) }{\t{\omega}_o} = \frac{\frac{C_s}{C_o}\t{E}_{n_i n_f}(S) }{\omega_o} =\frac{\frac{C_s}{C_o} \t{\omega}_s }{\omega_o} = \frac{\frac{C_s}{C_o} \frac{C_o}{C_s}\omega_s }{\omega_o} = \frac{\omega_s}{\omega_o} = (1+z).
\ee
We remark that this quantity is indeed frame invariant as expected for an observable.

The spacelike four-vectors normalised with respect to the metric $\t{g}_{\mu\nu}$ and pointing into the photon direction at the observer and at the source are $\t{n}_o = (0,0,0,\t{a}_o^{-1})$ and $\t{n}_s = (0,0,0,\t{a}_s^{-1})$.
From the equations of motion~\eqref{eq:A1} and \eqref{eq:A2}, one finds the solutions
\begin{align}
A_\sigma(\eta,r) = & \sum_{\lambda= \pm}\frac{\t{a}_s A_I^{(\lambda)}}{\t{r}_{phys}} \cos \l( \t{\omega}(r- \eta) + \t{\varphi}_I^{(\lambda)} \r) \epsilon_\sigma^{(\lambda)} \,, \\
A^\nu(\eta,r) = & \sum_{\lambda= \pm} \frac{\t{a}_s}{\t{a}^2(\eta)} \frac{A_I^{(\lambda)}}{\t{r}_{phys}} \cos \l( \t{\omega}( r- \eta ) + \t{\varphi}_I^{(\lambda)}\r) \epsilon^\nu_{(\lambda)} \,.
\end{align}
As in Sec.~\ref{sec:EMradiation}, the prefactor $\t{a}_s^2/\t{a}^2(\eta)$ introduced by the cosmological damping of $A^\mu(\eta,r)$ is split between $A^\mu$ and $A_\nu$ to preserve $A^\mu = \t{g}^{\mu\nu} A_\nu$
and the $\t{r}_{phys}$ in the denominators represent physical distances at the time of emission. The physical distances in the different frames, associated to the same comoving distance, are related by
\be \label{eq:rphys}
 \t{r}_{phys} = \frac{C}{C_o} r_{phys} = \frac{C}{C_o} \frac{a}{a_o} r = \frac{C}{C_o} a r \,,
\ee
where the normalisation is chosen such that comoving distances and physical distances match today. 
Finally, peforming the analogue calculation as in the Jordan frame to compute the flux Eq.~\eqref{eq:JordanFlux}, using the appropriately defined Einstein-frame quantities laid out here, we derive the Einstein-frame flux in the frame of the observer, finding 
\begin{align}
\t{F}_o = &-\l(\t{T}_{\mu\nu}^{em} \t{u}_o^\mu \t{n}^\nu \r)_o \\
= &-\l(\t{T}_{0z}^{em}\r)_o  \t{u}_o^0 \t{n}^z_o = \t{a}_o^{-2} \l( \frac{1}{\pi} \t{g}^{\alpha \beta} F_{0 \alpha} F_{z \beta} \r)_o =  -\t{a}_o^{-2}\l( \frac{1}{\pi} \p_0 A_\alpha \p_z A^\alpha \r)_o \\ = & -\t{a}_o^{-2} \Bigg[ \frac{1}{\pi} \frac{\dd}{\dd \eta} \l( \sum_{\lambda =\pm} \frac{\t{a}_s A_I^{(\lambda)}}{\t{r}_{phys}} \cos \l(\t{\omega}(r - \eta) + \t{\varphi}_I^{(\lambda)} \r) \epsilon_\alpha^{(\lambda)} \r) \\
& \cdot \underbrace{\frac{\dd r}{\dd z}}_{=1} \frac{\dd}{\dd r} \l( \sum_{\lambda' = \pm }  \frac{\t{a}_s}{\t{a}^2(\eta)} \frac{A_I^{(\lambda')}}{\t{r}_{phys}}  \cos \l( \t{\omega}(r - \eta) + \t{\varphi}_I^{(\lambda')}\r) \epsilon^\alpha_{(\lambda')} \r)\Bigg]_o\\
= &  \frac{1}{\pi \t{a}_o^2}  \Bigg[ \sum_{\lambda = \pm} \sum_{\lambda'= \pm}\epsilon^{(\lambda)}_\alpha \epsilon_{(\lambda')}^\alpha  \frac{A_I^{(\lambda)}A_I^{(\lambda')} \t{a}_s^2}{\frac{r^2 \t{a}_s^2}{C_o^2} \t{a}^2} \t{\omega}^2  \\ & \cdot \sin \l(\t{\omega}(r - \eta) + \t{\varphi}_I^{(\lambda)} \r)\sin \l(\t{\omega}(r - \eta) + \t{\varphi}_I^{(\lambda')} \r) \Bigg]_o + \mathcal{O}(r^{-3})\\
= &  \frac{1}{\pi a_o^2}  \l[ \frac{\t{\omega}^2 }{r^2 \t{a}^2}   \r]_o \l[\sum_{\lambda=\pm} \l(A_I^{(\lambda)} \sin \l( \t{\omega}(r - \eta) + \t{\varphi}_I^{(\lambda)} \r) \r)^2 \r]+ \mathcal{O}(r^{-3})\\
\simeq & \frac{1}{4 \pi r^2 a_o^4 C_o^2} 4  C_o^2 \omega^2 \l[\sum_{\lambda=\pm} \l(A_I^{(\lambda)} \sin \l( C_o \omega(r - \eta) + \t{\varphi}_I^{(\lambda)} \r) \r)^2 \r] 
\end{align}
\begin{align}
= & \frac{1}{4 \pi r^2 a_o^4 } 4  \omega_s^2 a_s^2 \l[\sum_{\lambda=\pm} \l(A_I^{(\lambda)} \sin \l(  C_o a_s \omega_s(r - \eta) + \t{\varphi}_I^{(\lambda)} \r) \r)^2 \r]  \\
= & \frac{1}{4 \pi r^2 a_o^2 (1+z)^2 } 4  \omega_s^2  \l[\sum_{\lambda=\pm} \l(A_I^{(\lambda)} \sin \l( C_o \omega(r - \eta) + \t{\varphi}_I^{(\lambda)} \r) \r)^2 \r]\,.
\label{eq:EFflux}
\end{align}
Eq.~\eqref{eq:EFflux} defines the Einstein-frame luminosity
\be
 \t{\mathcal{L}}_s= 4  \omega_s^2  \l[\sum_{\lambda=\pm} \l(A_I^{(\lambda)} \sin \l( C_o \omega(r - \eta) + \t{\varphi}_I^{(\lambda)} \r) \r)^2 \r] \,,
\ee
which upon averaging over time (see Eq. \eqref{SinIntegral}) yields the same result as in Eq.~\eqref{eq:JordanFrameL} up to some negligible terms, and the Einstein-frame luminosity distance
\be
\t{d}_L^{em} = r a_o(1+z) \,,
\ee
which is indeed frame invariant as can be seen from Eq.~\eqref{eq:JordanFrameDL}. It should be noted that this is a definition that maintains the frame invariance of the flux. Any other definition for which the luminosity and the squared luminosity distance transform in the same way would be acceptable since only the ratio of the two is observable.

Note that the distance inferred from supernovae Type Ia is also frame invariant since in the Einstein frame, $\t{G}_s =\t{G}_o$ and the nucleon mass $\t{m}_N(S) = C_o C_s^{-1} \t{m}_N(O)$ in Eq. \eqref{eq:ChandrasekharMass}, yielding
\be
\t{M}_{Ch}(S) = \l(\frac{C_s}{C_o}\r)^2 \t{M}_{Ch}(O) \,.
\ee
This gives the same $C$ dependence for the supernovae Type Ia luminosity distance
\be
\t{d}_L^{\tiny{\hbox{SNIa}}} = \l(\frac{C_o}{C_s}\r)^\gamma \t{d}_L^{em} =  \l(\frac{C_o}{C_s}\r)^\gamma d_L^{em} = d_L^{\tiny{\hbox{SNIa}}} \,,
\ee
which confirms the frame-invariance of the luminosity distance inferred from supernovae Type Ia (see Eq. \eqref{eq:JFdLSNIa}).

\subsection{Gravitational waves} \label{sec:EframeGW}

Recall that in the Einstein frame, one recovers the standard Einstein field equations of GR from the variation of the Einstein-Hilbert action, and the wave equation is given by (Sec.~\ref{sec:EFEoMs})
\be \label{eq:EinsteinFrameEoM}
 \p_\eta^2 \t{h}_{ij} + 2\t{\mathcal{H}} \p_\eta \t{h}_{ij} - \vec{\nabla}^2 \t{h}_{ij} =0
\ee
for the tensor perturbations $\t{h}_{ij}(\eta,\mathbf{x})$ around the flat FLRW Einstein frame metric $\t{g}_{ij}(\eta,\mathbf{x})= \t{a}^2(\eta)(\delta_{ij} + \t{h}_{ij}(\eta,\mathbf{x}))$, where $\mathcal{H}= \frac{\t{a}'}{\t{a}}$.
Note that we are still studying the case of a uniform background scalar field $C=C(\t{t})$, which implies $\t{a}(\t{t})= C(\t{t})a(\t{t})$, and therefore we will use $C_o = C(\t{t}_o) = C(0)$.
The two polarisations of the gravitational wave signal emitted by the binary system in the TT gauge are to lowest PN order~\cite{Maggiore:1900zz}:
\begin{align}
\t{h}_+(\t{\tau}_s) = & \frac{4 (\t{G}_{\tiny{\hbox{eff}}}(S) \t{M}_c(S))^{5/3}}{\t{r}_{phys}(S)} \l( \pi \t{f}_s \r)^{2/3} \frac{1 + \cos^2 \imath}{2} \cos \l( \t{\Phi}_s(\t{\tau}_s,M_c)\r) \,, \\
\t{h}_\times(\t{\tau}_s) = &  \frac{4 (\t{G}_{\tiny{\hbox{eff}}}(S) \t{M}_c(S))^{5/3}}{\t{r}_{phys}(S)} \l( \pi \t{f}_s \r)^{2/3} \cos (\imath) \sin \l( \t{\Phi}_s(\t{\tau}_s, M_c) \r) \,,
\end{align}
where $\t{r}_{phys}(S) = \frac{C_s}{C_o} a_s r$ was defined in Eq.~\eqref{eq:rphys}.
Since we are in the Einstein frame, the Newton gravitational constant is unchanged, $\t{G}_{\tiny{\hbox{eff}}}= G_N = \hbox{cst}$, whereas the masses are $C$ dependent (see Sec.~\ref{sec:spinors}).
Specifically,
\be
 \t{M_c}(S)= \frac{C_o}{C_s} \t{M}_c(O) = \frac{C_o}{C_s} M_c(O) \,,
\ee
where we normalised the masses today $\t{M}_c(O) = M_c(O)$.
The gravitational wave frequency is twice the orbital frequency and we have for any of the two
\be
 \frac{\t{\omega}_s}{\t{\omega}_o} = \frac{C_o}{C_s} \frac{\omega_s}{\omega_o} = \frac{C_o}{C_s} (1+z) \,, \label{eq:falseredshift}
\ee
following the same argumentation as in Eq.~\eqref{eq:unscreened_frequencies}.
Recall that Eq.~\eqref{eq:falseredshift} does not imply a frame dependence of the measured redshift (Sec.~\ref{sec:EFFlux}). We now wish to determine the measured mass.
Remembering the transformation law of the spacetime invariant (Eq.~\eqref{eq:spacetimeinvariant}), proper time transforms as
\be
 \dd\t{\tau}_s = \frac{\dd \t{\tau}_o}{(1+z) \frac{C_o}{C_s}}
\ee
and the time variation of the frequency in the source frame is given by~\cite{Maggiore:1900zz}
\begin{align}
\frac{\dd \t{f}_{gw}^{(s)}}{\dd \t{\tau}_s} =  &\frac{96}{5} \pi^{8/3} \l( \t{G}_{\tiny{\hbox{eff}}}(S) \t{M}_c(S) \r)^{5/3} \l[ \t{f}_{gw}^{(s)}\r]^{11/3}
\end{align}
such that in the frame of the observer,
\begin{align}
\frac{\dd \t{f}_{gw}^{(o)}}{\dd \t{\tau}_o} 
=  &\frac{96}{5} \pi^{8/3} \l( \t{G}_{\tiny{\hbox{eff}}}(O) \mathcal{M}_c(z) \l(\frac{C_o}{C_s} \r)^2\r)^{5/3} \l[ \t{f}_{gw}^{(o)}\r]^{11/3} \,.
\end{align}
Since measuring $\t{f}_{gw}^{(O)}(\t{\tau}_o)$, one again measures the effective redshifted chirp mass (see Eq.~\eqref{eq:EffectiveChirpMass}),
\be
 \mathcal{M}_{\tiny{\hbox{c,eff}}}(z)\equiv \l( \frac{C_o}{C_s}\r)^2 \mathcal{M}_c(z) = \l( \frac{C_o}{C_s}\r)^2 (1+z)  M_c(O) \,.
\ee
Next, let us inspect the transformation of $\t{\Phi}_s(\t{\tau}_s)$.
In the Einstein frame,
\begin{align}
 \t{\Phi}(\t{\tau}_o,\mathcal{M}_{\tiny{\hbox{c,eff}}}(z) ) =  -2 \l( 5 \t{G}_{\tiny{\hbox{eff}}}(O)  \mathcal{M}_{\tiny{\hbox{c,eff}}}(z) \r)^{-5/8} \t{\tau}_o^{5/8} + \Phi_o \,,
\end{align}
which corresponds to Eq.~\eqref{eq:JordanFramePhase}. Finally, from the equation of motion~\eqref{eq:EinsteinFrameEoM}, we find that the amplitude decreases as $1/(\t{a}(\eta)r)$ and therefore
\begin{align}
\t{h}_+(\t{\tau}_s) = & \frac{\t{a}_s}{\t{a}_o}\frac{4 (\t{G}_{\tiny{\hbox{eff}}}(S) \t{M}_c(S))^{5/3}}{\t{r}_{phys}} \l( \pi \t{f}_s \r)^{2/3} \frac{1 + \cos^2 \imath}{2} \cos \l( \t{\Phi}_s(\t{\tau}_s, \t{M}_c(S))\r) \\
= & \frac{\t{a}_s}{\t{a}_o}\frac{4 \l(G_N \t{M}_c(O)\frac{\t{M}_c(S)}{\t{M}_c(O)} \r)^{5/3}}{r  \frac{\t{a}_s}{C_o}} \l( \pi \t{f}_s \r)^{2/3} \frac{1 + \cos^2 \imath}{2} \cos \l( \t{\Phi}_s(\t{\tau}_s, \t{M}_c(S))\r) \\
= & \frac{4 \l(G_N M_c(O)\frac{C_o}{C_s} \r)^{5/3}}{r a_o} \l( \pi \t{f}_o \r)^{2/3}(1+z)^{2/3} \l(\frac{C_o}{C_s} \r)^{2/3} \frac{1 + \cos^2 \imath}{2} \cos \l( \t{\Phi}_o(\t{\tau}_o, \mathcal{M}_{\tiny{\hbox{c,eff}}}(z) )\r) \\
= & \frac{4 \l(G_N M_c(O)(1+z) \l(\frac{C_o}{C_s}\r)^2 \r)^{5/3}}{r a_o(1+z) \frac{C_o}{C_s}}  \l( \pi \t{f}_o \r)^{2/3} \frac{1 + \cos^2 \imath}{2} \cos \l( \t{\Phi}_o(\t{\tau}_o, \mathcal{M}_{\tiny{\hbox{c,eff}}}(z) )\r) \\
= & \frac{4 \l(G_N \mathcal{M}_{\tiny{\hbox{c,eff}}}(z) \r)^{5/3}}{r a_o(1+z) \frac{C_o}{C_s}}  \l( \pi \t{f}_o \r)^{2/3} \frac{1 + \cos^2 \imath}{2} \cos \l( \t{\Phi}_o(\t{\tau}_o, \mathcal{M}_{\tiny{\hbox{c,eff}}}(z) )\r) \,, \\
\end{align}
where we have used that $\t{\Phi}_s(\t{\tau}_s, \t{M}_c(S)) = \t{\Phi}_o(\t{\tau}_o, \mathcal{M}_{\tiny{\hbox{c,eff}}}(z))$.
Thus, we find that
\be
 \t{d}_L^{gw} = r a_o (1+ z) \frac{C_o}{C_s} = d_L^{gw} \,,
\ee
confirming the frame invariance of the gravitational wave luminosity distance.

\section{The impact of screening mechanisms} \label{sec:screeningmechanism}

We have derived the luminosity distances that are inferred from gravitational waves and light in the Jordan and Einstein frames and confirmed their frame invariant difference in the presence of a Planck mass evolution.
We shall now address the question of whether this Planck mass evolution should be interpreted as that in the cosmological background at the time of emission and observation, or if instead this evolution should be considered locally at the source and observer, and hence should be screened to comply with stringent astrophysical tests of GR. For this purpose, we will take an abstract approach to screening and assume that in a bubble $\mathcal{B} \subset \mathcal{M}$ around the source and the observer, the effective Planck mass reduces to the bare Planck mass
\be
M^2(t,\mathbf{x})= C^2(t, \mathbf{x}) M_p^2 =M_p^2, ~~ \forall (t,\mathbf{x}) \in \mathcal{B} \label{eq:screeningeffect}
\,.
\ee
We will furthermore assume that the bubble does not have a sharp edge.
There should not be a sudden step but a continuous process where the effective Planck mass smoothly interpolates between the background value outside the high-density region and that inside.
The scale of variation of the effective Planck mass should be large enough to pass Solar-System tests of gravity~\cite{Will:2014kxa} and in particular the lunar laser-ranging constraints~\cite{Williams:2004qba}.
Hence, we will assume this scale to be larger than the size of the Solar System.
Note that such a scenario can, for instance, be realised with the chameleon screening mechanism~\cite{Lombriser:2013eza}, in which case the background value of the chameleon field would lie well below the sensitivity of Standard Sirens tests~\cite{Lombriser:2015sxa,Belgacem:2019pkk}.
Our approach shall, however, be agnostic about whether a more effective viable screening effect could be realised in the general Horndeski action or not.
As pointed out in Secs.~\ref{sec:intro} and \ref{sec:standardsirens}, we will find that Standard Sirens test the local couplings, which are either not screened, in which case one must adopt stringent astrophysical constraints on the local evolving coupling that are several orders of magnitude stronger than the sensitivity of Standard Sirens and leave no detectable signature for those.
Or the local couplings are screened, in which case, as in GR, the luminosity distances inferred from Standard Sirens will agree with the electromagnetic ones.
Hence, our conclusions will not be affected by the particulars of the implemented screening mechanism, and we may safely operate with our abstract notion of screening.

\subsection{Screening in the Jordan frame} \label{sec:JFscreening}

The computations in Jordan frame performed for light in Sec.~\ref{sec:JordanFrame} and the notion of redshift are not affected by the screening effect imposed by Eq.~\eqref{eq:screeningeffect}, except for its impact on the production mechanism of supernovae Type Ia, which we will only briefly discuss.
Therefore, we will focus on the impact on the gravitational waves. The gravitational action in the Jordan frame contains the term
\be
S \supset \frac{M_p^2}{2} \int \mathrm{d}^4x \sqrt{-g} G_4(\phi(t,\mathbf{x})) R = \frac{M_p^2}{2} \int \mathrm{d}^4x \sqrt{-g} C^2(t,\mathbf{x}) R \,,
\ee
which is the term that gives rise to the modified propagation equation of  gravitational waves in Sec.~\ref{sec:JFGWs}.
Previously, we have assumed a uniform scalar field $\phi(t)$ and thus a uniform $C(t)$.
However, to comply with stringent astrophysical constraints on deviations from GR, we will need to recover $C=1$ in regions where GR has been confirmed to high degree such as the Solar System, and we now allow for an explicit $\mathbf{x}$ dependence in $C(t,\mathbf{x})$ to account for these spatial variations in $G_4$ and hence in $G_{\tiny{\hbox{eff}}}$. The quadratic effective action can then be written as
\be
 S^{(2)} = \frac{M_p^2}{2} \int \mathrm{d}t \mathrm{d}^3x a^3(t)C^2(t,\mathbf{x}) \l( \frac{1}{4} \dot{h}_{ij} \dot{h}_{kl} - \frac{\delta^{mn}}{4 a^2(t)} \p_m h_{ij} \p_n h_{kl}  \r)\delta^{lj} \delta^{ki} \,.
\ee
Variation with respect to $h_{ij}$ and its first derivatives gives the wave equation
\be
\ddot{h}_{ij} + 3 \frac{\dot{a}}{a} \dot{h}_{ij} + 2 \frac{\dot{C}}{C} \dot{h}_{ij} - \frac{\vec{\nabla}^2}{a^2}  h_{ij} - 2\frac{\delta^{kl}}{a^2} \frac{\p_k C}{C} \p_l h_{ij} = 0 \,,
\ee
which in conformal time $\mathrm{d}\eta = \frac{\mathrm{d}t}{a}$ becomes
\be
 h_{ij}''(\eta,\mathbf{x})+ 2 \l( \mathcal{H}(\eta) + \frac{C'(\eta, \mathbf{x})}{C(\eta,\mathbf{x})}\r) h_{ij}'(\eta,\mathbf{x}) -\vec{\nabla}^2 h_{ij}(\eta,\mathbf{x}) - 2 \delta^{kl}\frac{\p_k C(\eta,\mathbf{x})}{C(\eta,\mathbf{x})}\p_l h_{ij}(\eta,\mathbf{x}) =0 \,,
\ee
where we have kept track of the spacetime dependence for clarity.
To simplify the equation we define an effective Hubble rate in conformal time that is also space dependent
\be
 \t{\mathcal{H}}(\eta, \mathbf{x}) \equiv \frac{\t{a}'(\eta, \mathbf{x})}{\t{a}(\eta, \mathbf{x})} = \mathcal{H}(\eta) + \frac{C'(\eta, \mathbf{x})}{C(\eta, \mathbf{x})} \,.
\ee
Since both polarisations satisfy the same equation of motion, we will again denote both by $h_A$ and we obtain
\be \label{eq:ScreenedEoM}
h_A''(\eta,\mathbf{x})+ 2 \mathcal{\t{H}}(\eta, \mathbf{x}) h_A'(\eta,\mathbf{x}) -\vec{\nabla}^2 h_A(\eta,\mathbf{x}) - 2 \delta^{ij}\frac{\p_i C(\eta,\mathbf{x})}{C(\eta,\mathbf{x})}\p_j h_A(\eta,\mathbf{x}) =0 \,.
\ee
Furthermore defining an auxiliary field $\chi = \chi (\eta, \mathbf{x})$ with
\be
 h_A(\eta,\mathbf{x}) = \frac{1}{a(\eta) C(\eta, \mathbf{x})} \chi(\eta,\mathbf{x}) = \frac{1}{\t{a}(\eta, \mathbf{x})} \chi(\eta,\mathbf{x}) \,.
\ee
Eq.~\eqref{eq:ScreenedEoM} becomes
\be
 \chi'' - \frac{\t{a}''}{\t{a}} \chi - \l(\vec{\nabla}^2 \chi + \frac{\vec{\nabla} C}{ C} \vec{\nabla} \chi - \frac{\vec{\nabla}^2 C}{C} \chi - \frac{(\vec{\nabla} C)^2}{C^2} \chi   \r) =0 \,.
\ee
We now wish to solve this equation and assume that the gravitational wave propagates radially on a sphere such that
\be
 \chi(\eta, r, \theta, \phi) = \frac{1}{r}f(\eta,r) \,.
\ee
We are interested in the limit where $r$ is large and may hence neglect variations of $C$ along $\theta$ and $\phi$ that scale as $\mathcal{O}(r^{-2})$ such that $C(\eta,r,\theta, \phi)=C(\eta, r)$.
Morphology dependent screening mechanisms such as the Vainshtein mechanism~\cite{Falck:2014jwa,Bloomfield:2014zfa} can thus be treated in the same manner, where we however note that the Vainshtein mechanism may not screen the time variation of the background value of the scalar field.
We get
\be
f'' - \frac{\t{a}''}{\t{a}} f - \l( \p_r^2 f + \frac{\p_r C}{C} \l( \p_r f - \frac{f}{r}\r) - \frac{\p_r^2C}{C}f - \frac{2\p_r C}{r C} f - \frac{\l(\p_r C\r)^2}{C}f \r) =0 \label{eq:screendedwave}
\ee
and further neglect terms that scale as $1/r$. Furthermore, we may approximate $\p_r C \sim C r^{-1} \rightarrow 0$. More specifically, multiplying Eq.~\eqref{eq:screendedwave} by $C^2$ and taking the Fourier transform yields
\begin{align} \label{eq:ScreenedGWapprox}
0 = \int \frac{\dd k}{(2\pi)}\frac{\dd k_1}{(2\pi)}\frac{\dd k_2}{(2\pi)} \Bigg( C(\eta,k_1) C(\eta,k_2) f''(\eta,k) - \frac{(a(\eta)C(\eta, k_1))''}{a(\eta)} C(\eta, k_2) f(\eta,k) \\- \l(- k^2 - k_1 k + k_1^2 + k_1 k_2 \r) C(\eta,k_1) C(\eta,k_2) f(\eta,k)\Bigg)e^{i (k+k_1+k_2) \cdot r}\,,
\end{align}
\begin{figure}
\centering
\includegraphics[scale=0.8]{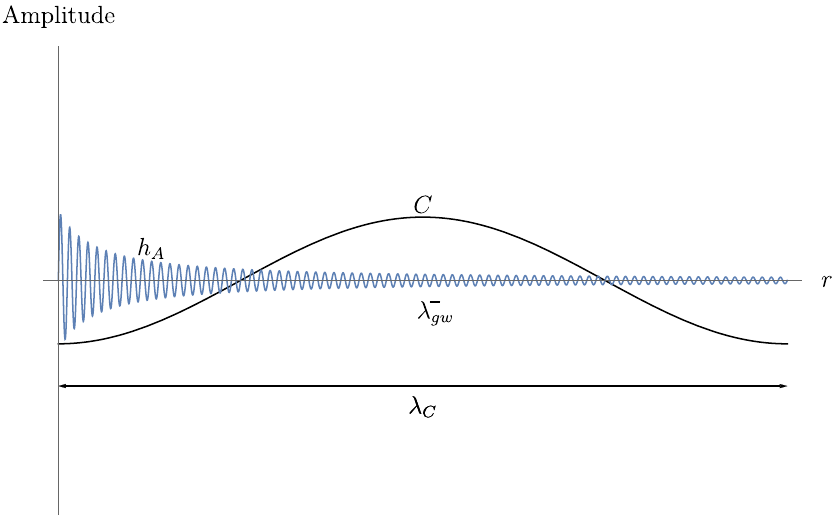}
\caption{Schematic representation of a gravitational wave propagation through a medium against the Planck mass evolution causing its friction.
Spatial variations of the field $C$ at a scale of $\lambda_C$ can be neglected on the scale set by the gravitational wavelength $\lambda_{gw}$ ($\lambda_C \gg \lambda_{gw}$).
This approximation will remain justified for the foreseeable gravitational wave experiments with $\lambda_{gw}$ at the order of a few kilometers whereas due to screening the effective Plank mass must vary on scales larger than the Solar System.}
\label{fig:approximationscreening}
\end{figure}
where $k_1$ and $k_2$ are the wavevectors of $C$, which are of the order of the scalar field wavevector $k_\phi$, and $k$ is the gravitational wavevector. As schematically represented in Fig. \ref{fig:approximationscreening}, the typical wavelength of $C$ is much larger than the gravitational wavelength $\lambda_c \simeq \lambda_\phi \gg \lambda_{gw}$. Since $k = \frac{2\pi}{\lambda}$, the Fourier coefficients $C(\eta,k_{1})$ and $C(\eta,k_{2})$ contribute most at their corresponding wavelength and are negligible whenever $k_1$ or $k_2$ is of the order of $k$. Hence, we can neglect all the terms that contain $k_{1}$ or $k_{2}$ next to $k^2$.
Thus, the only terms remaining are the ones without spatial derivatives with respect to $C$.
Keeping only the important terms in real space, this implies
\be
\p_\eta^2 f(\eta, r) - \frac{\t{a}''(\eta, r)}{\t{a} (\eta, r)} f (\eta, r)- \p_r^2 f (\eta,r)=0 
\,.
\ee
and after performing a Fourier transformation,
\be
f''(\eta,k ) + \l(k^2 - \frac{\t{a}''}{\t{a}} \r) f(\eta,k) =0 \,,
\ee
which for $k^2 \gg \frac{\t{a}''}{\t{a}} = \frac{a''}{a} + 2 \frac{a' C'}{aC} + \frac{C''}{C}$ yields
\be
f''(\eta, k) + k^2 f(\eta, k) =0 \,.
\ee
Hence, in contrast to the situation in standard GR, the amplitude of the wave decays as $1/(\t{a}(\eta,r)r)$ instead of $1/(a(\eta)r)$.
For $h_+$ for example, this implies that
\begin{align}
h_+ (\tau_s) = & \frac{\t{a}(\tau_s,r_s)}{\t{a}(\tau_o,r_o)}  \frac{4 (G_{\tiny{\hbox{eff}}}(S) M_c)^{5/3}}{a(\tau_s)r} \l( \pi f_s\r)^{2/3} \frac{1 + \cos^2 \imath}{2} \cos \l( \Phi_s(\tau_s, M_c)\r) \\
= & \frac{4 (G_N \mathcal{M}_c(z))^{5/3}}{d_L^{em}\frac{C(\tau_o,r_o)}{C(\tau_s,r_s)}} \l( \pi f_o\r)^{2/3} \frac{1 + \cos^2 \imath}{2} \cos \l( \Phi_s(\tau_s, M_c)\r) \\
=&  \frac{4 (G_N \mathcal{M}_c(z))^{5/3}}{d_L^{gw}} \l( \pi f_o\r)^{2/3} \frac{1 + \cos^2 \imath}{2} \cos \l( \Phi_o(\tau_o, \mathcal{M}_c(z))\r) \,,
\end{align}
where we set $G_{\tiny{\hbox{eff}}}(S)=G_N$ assuming screening at the source. We have also identified the gravitational wave luminosity distance
\be
 d_L^{gw} = r a_o (1+z) \frac{C(\tau_o,r_o)}{C(\tau_s,r_s)} \,.
\ee
which hence depends on the effective Planck masses at the locations of the source and of the observer.
If there exists an efficient screening mechanism at the source and at the observer with $C(\tau_o,r_o) = C(\tau_s,r_s) =1$ as required for observational compatibility with stringent astrophysical tests, the gravitational wave luminosity distance therefore reduces to that of standard GR,
\be \label{eq:JordanFrameScreenedDL}
\l(d_L^{gw} \r)_{\tiny{\hbox{screened}}}= r a_o (1+z),
\ee
and does not leave any signature of cosmological modifications of gravity in Standard Sirens.

Note also that in case of screening, the production mechanism of supernovae Type Ia is that of GR, in which these are standard candles. Since there is no difference with respect to GR in the production mechanism nor in the propagation of light, the inferred luminosity distance from supernovae Type Ia also reduces to
\be
\l(d_L^{\tiny{\hbox{SNIa}}}\r)_{\tiny{\hbox{screened}}}= ra_o(1 +z) = \l(d_L^{gw} \r)_{\tiny{\hbox{screened}}}\,.
\ee

\subsection{Screening in the Einstein frame} \label{sec:EFscreening}

We shall now inspect whether we find consistent results to Sec.~\ref{sec:JFscreening} in the Einstein frame.
We first examine the redshift in the presence of a screening mechanism (Sec.~\ref{sec:scEFredshift}) to then rederive the electromagnetic flux and luminosity distance (Sec.~\ref{sec:scEFflux}). Finally, in Sec.~\ref{sec:scEFGWs} we recompute the gravitational wave luminosity distance, confirming its equivalence to the electromagnetic luminosity distance in the presence of screening found for the Jordan frame in Sec.~\ref{sec:JFscreening}.

\subsubsection{Redshift} \label{sec:scEFredshift}

Recall the definition of redshift in Eq.~\eqref{eq:redshift}.
It depends on the four-velocities of the source and observer as well as on the relation between the value of the Bohr radius at the source and at the observer.
In a scenario where the source and observer are screened,
we have
\begin{align}
 &\t{E}_{n_in_f}(t,\mathbf{x}) \underbrace{C(t,\mathbf{x})}_{=1} =  \t{E}_{n_in_f}(t,\mathbf{x}) = E_{n_i n_f}, \forall (t,\mathbf{x}) \in \mathcal{B} \,, \\
 &\t{u}^\mu = \underbrace{C^{-1}(t,\mathbf{x})}_{=1} u^\mu = u^\mu , \forall (t,\mathbf{x}) \in \mathcal{B} \label{eq:screenedvelocities}
\end{align}
and, hence,
\be
 (1+\t{z}) = \frac{\t{\omega}_s}{\t{\omega}_o} = \frac{\omega_s}{\omega_o} = (1+z) \,.
\ee
The redshift is therefore frame invariant and independent of screening. Note that in order to find this consistency, we carefully took into account screening of the Bohr radius and of the four-velocities.

\subsubsection{Flux} \label{sec:scEFflux}

We will now inspect how the Einstein-frame flux is affected by screening. We again assume a screened source and observer, and thus their proper times remain unaffected by the conformal transformation.
Following Eq.~\eqref{eq:screenedvelocities}, their four-velocities are gvien by
\begin{align}
\t{u}_o^\mu = & \underbrace{C^{-1}(\eta_o, \mathbf{x}_o)}_{=1}  u_o^\mu=  (a_o^{-1},0,0,0) 
\,,\\
\t{u}_s^\mu = & \underbrace{C^{-1}(\eta_s, \mathbf{x}_s)}_{=1} u_s^\mu=  (a_s^{-1},0,0,0) \,,
\end{align}
which are normalised with respect to the screened metric $\t{g}_{\mu\nu}(S)= C^2(S) g_{\mu\nu}(S) = g_{\mu\nu}(S)$.

Since the electromagnetic action is conformally invariant, one derives the same electromagnetic equations of motion and one can easily show that photons follow null geodesic (see Sec~\ref{sec:optics}).
The geodesic equation implies
\be
 \frac{\dd \t{k}_\mu}{\dd \t{\lambda}} = \frac{1}{2} \l(\t{g}_{\alpha \kappa, \mu}\r) \t{k}^\alpha \t{k}^\kappa
\ee
and for the $\mu =i$ coordinate, we find
\be
 \frac{\dd \t{k}_i}{\dd \t{\lambda}} = \frac{1}{2} \l(\t{g}_{\alpha \kappa, i}\r) \t{k}^\alpha \t{k}^\kappa = a^2 C(x) (\p_i C(x)) \delta_{\alpha \kappa}\t{k}^\alpha \t{k}^\kappa \simeq 0 \,.
\ee
Since the metric is weakly dependent on the spatial coordinates, we must have $\t{k}_i = \hbox{cst}$ along geodesics, which also implies $\t{k}_0 = \hbox{cst}$ from $\t{k}_\mu \t{k}^\mu = 0$.
We then obtain
\begin{align}
 \t{k}_\nu(O) =& (-\t{\omega},0,0,\t{\omega}) \\
 \t{k}_\nu(S) =& (-\t{\omega},0,0, \t{\omega}) \,, \quad  \t{\omega} = \hbox{cst} \,.
\end{align}
It follows that
\be
 \t{k}_\mu = \t{g}_{\mu\nu} \t{k}^\nu = C^2 g_{\mu\nu} \frac{\dd x^\nu}{\dd \lambda} \frac{\dd \lambda}{\dd \t{\lambda}} = C^2 \frac{\dd \lambda}{\dd \t{\lambda}} k_\mu =  k_\mu
\ee
such that $\t{\omega} = \omega$, where we chose the affine parameter such that $\dd\t{\lambda} = C^2 \dd \lambda$ with $C^2$ absorbing the apparent fifth force on photons (see Sec.~\ref{sec:EFgeodesics}).
This preserves the null geodesic equation for photons in the Einstein frame and we find
\begin{align} \label{eq:screened_frequencies}
-\t{\omega}_s = & (\t{k} \cdot \t{u})_s = - \frac{\t{\omega}}{a_s} = - \omega_s \,, \\
-\t{\omega}_o = & (\t{k} \cdot \t{u})_o = - \frac{\t{\omega}}{a_o} = - \omega_o \,,
\end{align}
where we have used that the four-velocities are screened at the observer and at the source.
We therefore find
\be
\frac{\t{\omega}_s}{\t{\omega}_o} = \frac{\omega_s}{\omega_o} =(1+z) \,.
\ee
The spacelike four-vectors pointing into the photon direction at the observer are $\t{n}_o = (0,0,0, a_o^{-1})$ and $\t{n}_s = (0,0,0, a_s^{-1})$, which are normalised with respect to the \emph{screened} metric $g_{\mu\nu}$.
The equations of motion for the $A_\mu$ and $A^\nu$ fields are given by
\begin{align}
 &\p_\eta^2 A_\mu - \vec{\nabla}^2 A_\mu =0 \,, \label{eq:ScreenedEinsteinFrameEoM1} \\
 &\p_\eta^2 A^\nu + 4 \t{\mathcal{H}}(\eta, \mathbf{x})   \p_\eta A^\nu - \l( \mathbf{\vec{\nabla}}^2 - 2 \frac{\t{a}'' \t{a} + (\t{a}')^2}{\t{a}^2}\r) A^\nu + \mathcal{O}(\p_i \t{a}(x) A^\nu)=0 \,, \label{eq:ScreenedEinsteinFrameEoM2}
\end{align}
where $\t{\mathcal{H}}(\eta, \mathbf{x}) \equiv \frac{\p_\eta \t{a}(\eta, \mathbf{x})}{\t{a}(\eta, \mathbf{x})}$ and we have neglected all the terms containing spatial derivatives of $\t{a}$ as explained in the context of Eq.~\eqref{eq:ScreenedGWapprox}.
Analogously to the calculation of the gravitational waves in the Jordan frame for the presence of screening mechanisms, we derive the following solutions to Eqs.~\eqref{eq:ScreenedEinsteinFrameEoM1} and \eqref{eq:ScreenedEinsteinFrameEoM2}:
\begin{align}
A_\sigma(\eta,r) = & \sum_{\lambda= \pm}\frac{\t{a}(\eta_s, r_s) A_I^{(\lambda)}}{\t{r}_{phys}} \cos \l( \t{\omega}(r- \eta) + \t{\varphi}_I^{\lambda} \r) \epsilon_\sigma \,,\\
A^\nu(\eta,r) = & \sum_{\lambda=\pm}\frac{\t{a}(\eta_s, r_s)}{\t{a}^2(\eta, r)} \frac{A_I^{(\lambda)}}{\t{r}_{phys}} \cos \l( \t{\omega}( r- \eta ) + \t{\varphi}_I^{(\lambda)}\r) \epsilon^\nu \,,
\end{align}
where $\t{r}_{phys}$ in the denominators represents physical distance at the time of emission. In screened regions of the Einstein frame
\be
 \t{r}_{phys}(S) = \underbrace{C(\eta_s, \mathbf{x}_s)}_{=1} r_{phys}  =a_s r \,,
\ee
where the normalisation is chosen such that comoving distances and physical distances agree today. The flux in the presence of screening in the Einstein frame becomes
\begin{align}
\t{F}_o 
= &-\l(\t{T}_{\mu\nu}^{em} \t{u}_o^\mu \t{n}^\nu \r)_o = -\l(\t{T}_{0z}^{em} \r)_o \t{u}_o^0 \t{n}_o^z  = a_o^{-2} \l( \frac{1}{\pi} F_{0 \alpha} \te{F}{_z ^\alpha}  \r)_o = -a_o^{-2}\l( \frac{1}{\pi} \p_0 A_\alpha \p_z A^\alpha \r)_o \\
= &-a_o^{-2} \Bigg[ \frac{1}{\pi} \frac{\dd}{\dd \eta} \l(  \sum_{\lambda =\pm} \frac{\t{a}_s A_I^{(\lambda)}}{\t{r}_{phys}} \cos \l(\t{\omega}(r - \eta) + \t{\varphi}_I^{(\lambda)}\r) \epsilon_\alpha^{(\lambda)} \r) \\
& \cdot \underbrace{\frac{\dd r}{\dd z}}_{=1} \frac{\dd}{\dd r} \l( \sum_{\lambda' = \pm }  \frac{\t{a}_s}{\t{a}^2(\eta,r)} \frac{A_I^{(\lambda')}}{\t{r}_{phys}}  \cos \l( \t{\omega}(r - \eta) + \t{\varphi}_I^{(\lambda')}\r) \epsilon^\alpha_{(\lambda')} \r)\Bigg]_o \\
= & \frac{1}{\pi  a_o^{2}}  \Bigg[ \sum_{\lambda = \pm} \sum_{\lambda'= \pm}\epsilon^{(\lambda)}_\alpha \epsilon_{(\lambda')}^\alpha   \frac{A_I^{(\lambda)}A_I^{(\lambda')} \t{a}_s^2}{ \t{a}^2 r^2 a_s ^2} \t{\omega}^2  \\ &\cdot \sin \l(\t{\omega}(r - \eta) + \t{\varphi}_I^{(\lambda)}\r) \sin \l(\t{\omega}(r - \eta) + \t{\varphi}_I^{(\lambda')}\r) \Bigg]_o + \mathcal{O}(r^{-3})
\end{align}
\begin{align}
= &  \frac{1}{\pi a_o^2}  \l[ \frac{ C_s^2 }{r^2 \t{a}^2} \t{\omega}^2 \r]_o \l[\sum_{\lambda=\pm} \l(A_I^{(\lambda)} \sin \l( \t{\omega}(r- \eta) + \t{\varphi}_I^{(\lambda)}\r)\r)^2 \r]+ \mathcal{O}(r^{-3}) \\
\simeq & \frac{1}{4 \pi r^2 a_o^4 C_o^2} 4 C_s^2  \t{\omega}^2  \l[\sum_{\lambda=\pm} \l(A_I^{(\lambda)} \sin \l(  \omega(r- \eta) + \t{\varphi}_I^{(\lambda)}\r)\r)^2 \r] \\
= & \frac{1}{4 \pi r^2 a_o^4 C_o^2  } 4 C_s^2 \omega_s^2 a_s^2 \l[\sum_{\lambda=\pm} \l(A_I^{(\lambda)} \sin \l(  a_s \omega_s(r- \eta) + \t{\varphi}_I^{(\lambda)}\r)\r)^2 \r]  \\
= & \frac{1}{4 \pi r^2 a_o^2 (1+z)^2 \l(\frac{C_o}{C_s} \r)^2} 4 \omega_s^2 \l[\sum_{\lambda=\pm} \l(A_I^{(\lambda)} \sin \l(  a_s \omega_s(r- \eta) + \t{\varphi}_I^{(\lambda)}\r)\r)^2 \r]\,.
\end{align}
The luminosity distance in the presence of screening therefore is
\be
 (\t{d}_L^{em})_{\tiny{\hbox{screened}}} = r  a_o (1+z) \frac{C(\eta_o,r_o)}{C(\eta_s, r_s)} \,,
\ee
where the screened conformal factors $C_o = C(\eta_o, r_o) = 1$ and $C_s = C(\eta_s, r_s) =1$ are introduced by evaluating $\t{a}(\eta, r)$ at the observer and at the source.
Hence, when screening operates,
\be
 (\t{d}_L^{em})_{\tiny{\hbox{screened}}} = r  a_o (1+z) \,.
\ee
As expected, the electromagnetic luminosity distance therefore does not change in the presence of a screening mechanism.

The production mechanism of supernovae Type Ia is that of GR in the case of screening, where these are standard candles (see Sec.~\ref{sec:optics}), leading to
\be
\l(\t{d}_L^{\tiny{\hbox{SNIa}}}\r)_{\tiny{\hbox{screened}}} = r a_o (1+z)\,.
\ee
The luminosity distance inferred from supernovae Type Ia in the screened scenario hence also agrees with GR in the Einstein frame.

\subsubsection{Gravitational waves} \label{sec:scEFGWs}

Finally, let us determine the effect on gravitational waves from the presence of screening in the Einstein frame.
The metric satisfies here the standard Einstein field equations obtained from the Einstein-Hilbert action. Consider the tensor perturbations $\t{h}_{ij}(\eta,\mathbf{x}) = h_{ij} (\eta,\mathbf{x})$ around the flat FLRW Einstein-frame metric $\t{g}_{ij}(\eta,\mathbf{x})= \t{a}^2(\eta, \mathbf{x})(\delta_{ij} + \t{h}_{ij}(\eta,\mathbf{x}))$.
In the absence of anisotropic stress, these must satisfy the wave equation in curved spacetime,
\be
 \t{\nabla}^\rho \t{\nabla}_\rho \t{h}_{\mu\nu} + 2 \t{R}_{\mu \rho \nu \sigma} \t{h}^{\rho \sigma} =0 \,.
\ee
Note, however, that a small contribution of anisotropic stress will be introduced by neutrinos~\cite{Weinberg:2003ur}.
Evaluating the covariant derivatives and the Riemann tensor adopting the flat FLRW metric and accounting for the spatial dependence of $C(t,\mathbf{x})$ in the metric, we obtain the wave equation
\be\label{EinsteinFrameEoM}
\p_\eta^2 \t{h}_{ij} + \t{\mathcal{H}}(\eta, \mathbf{x}) \p_\eta \t{h}_{ij} - \mathbf{\nabla}^2 \t{h}_{ij} + \mathcal{O}((\p_k \t{a})\t{h}_{ij})=0 \,,
\ee
where $\t{\mathcal{H}}(\eta, \mathbf{x}) = \frac{\p_\eta \t{a}(\eta, \mathbf{x})}{\t{a}(\eta, \mathbf{x})}$.
We have again neglected all the terms containing spatial derivatives of $\t{a}(\eta,\mathbf{x})$ as $\lambda_c \gg \lambda_{gw}$ (see Sec.~\ref{sec:JFscreening} and Fig.~\ref{fig:approximationscreening}). 

The two polarisations of the gravitational wave signal emitted by the binary in the TT gauge are
\begin{align}
 \t{h}_+(\t{\tau}_s) = & \frac{4 (\t{G}_{\tiny{\hbox{eff}}}(S) \t{M}_c(S))^{5/3}}{\t{r}_{phys}(S)} \l( \pi \t{f}_s \r)^{2/3} \frac{1 + \cos^2 \imath}{2} \cos \l( \t{\Phi}_s(\t{\tau}_s, \t{M}_c(S))\r) \,, \\
 \t{h}_\times(\t{\tau}_s) = &  \frac{4 (\t{G}_{\tiny{\hbox{eff}}}(S) \t{M}_c(S))^{5/3}}{\t{r}_{phys}(S)} \l( \pi \t{f}_s \r)^{2/3} \cos (\imath) \sin \l( \t{\Phi}_s(\t{\tau}_s,\t{M}_c(S)) \r) \,,
\end{align}
where $\t{G}_{\tiny{\hbox{eff}}}(S)=G_N$ and $\t{M}_c(S)= \t{M}_c(O)$ because of screening.
In the local wave zone, the screened Einstein frame physical distance is $\t{r}_{phys}(S) = a_s r$ with the comoving coordinate distance $r$.
Following the same reasoning as for photons in  Eq.~\eqref{eq:screened_frequencies}, the gravitational wave frequencies are transformed from the rest frame of the source to that of the observer as
\be
 \frac{\t{\omega}_s}{\t{\omega}_o} =  \frac{\omega_s}{\omega_o} =  (1+z) \,.
\ee
To describe the effect on the chirp mass, we consider the screened proper time
\be
 \dd \t{\tau}_s =  \underbrace{C(\t{\tau_s}, \mathbf{x}_s)}_{=1}\dd \tau_s = \dd \tau_o (1+z)^{-1} = \frac{\dd \t{\tau}_o}{(1+z)}
\ee
such that
\begin{align}
 \frac{\dd \t{f}_{gw}^{(S)}}{\dd \t{\tau}_s} =  &\frac{96}{5} \pi^{8/3} \l( \t{G}_{\tiny{\hbox{eff}}}(S) \t{M}_c(S) \r)^{5/3} \l[ \t{f}_{gw}^{(S)}\r]^{11/3} \\
 =  &\frac{96}{5} \pi^{8/3} \l( G_N \t{M}_c(O) \r)^{5/3} \l[ \t{f}_{gw}^{(S)}\r]^{11/3}
\end{align}
and therefore,
\begin{align}
 \frac{\dd \t{f}_{gw}^{(O)}}{\dd \t{\tau}_o}   =  &\frac{96}{5} \pi^{8/3} \l( G_N \mathcal{\t{M}}_c(z) \r)^{5/3} \l[ \t{f}_{gw}^{(O)}\r]^{11/3} \,.
\end{align}
Since observing $\t{f}_{gw}^{(O)}(\t{\tau})$, in contrast to the unscreened scenario in Eq.~\eqref{eq:EffectiveChirpMass} we measure here the redshifted chirp mass, which is the same as in the Jordan frame and as in GR
\be
 \mathcal{\t{M}}_c(z) = (1+z) \t{M}_c(O)= \mathcal{M}_c(z) \,.
\ee
The phase transforms as:
\begin{align}
 \t{\Phi}(\t{\tau}_s) = & -2 \l( 5 \t{G}_{\tiny{\hbox{eff}}}(S) \t{M}_c(S)) \r)^{-5/8} \t{\tau}_s^{5/8} + \Phi_o \\
 =&  -2 \l( 5 G_N  \mathcal{\t{M}}_c(z) \r)^{-5/8} \t{\tau}_o^{5/8}+ \Phi_o \,.
\end{align}
For the amplitude, we find from the equation of motion~\eqref{EinsteinFrameEoM} a scaling of $1/(\t{a}(\eta,r )r)$ and therefore
\begin{align}
 \t{h}_+(\t{t}_o,r_o) = & \frac{\t{a}(\eta_s, r_s)}{\t{a}(\eta_o, r_o)}  \frac{4 (G_{\tiny{\hbox{eff}}}(S) \t{M}_c(S))^{5/3}}{\t{r}_{phys}} \l( \pi \t{f}_s \r)^{2/3} \frac{1 + \cos^2 \imath}{2} \cos \l( \t{\Phi}_s(\t{\tau}_s, \t{M}_c(S))\r) \\
 = & \frac{4 \l(G_N \t{\mathcal{M}}_c(z)  \r)^{5/3}}{ra_o(1+z) \frac{C(\eta_o, r_o)}{C(\eta_s, r_s)}}  \l( \pi \t{f}_o \r)^{2/3} \frac{1 + \cos^2 \imath}{2} \cos \l( \t{\Phi}_o(\t{\tau}_o, \t{\mathcal{M}}_c(z))\r) \,,
\end{align}
where we have used that $\t{\Phi}_s(\t{\tau}_s, \t{M}_c(S)) = \t{\Phi}_o(\t{\tau}_o, \t{\mathcal{M}}_c(z))$.
We identify the gravitational wave luminosity distance
\be
\t{d}_L^{gw} = r a_o (1+ z) \frac{C(\eta_o, r_o)}{C(\eta_s, r_s)} \,.
\ee
$C(\eta_o, r_o) $ and $C(\eta_s, r_s)$ again arise from evaluating $\t{a}$ at the source and at the observer, taken to be screened (i.e., $C(\eta_o, r_o) =1 =C(\eta_s, r_s) $), and therefore,
\be
\l(\t{d}_L^{gw}\r)_{\tiny{\hbox{screened}}} = r a_o (1+ z) = d_L^{em} = \l(d_L^{\tiny{\hbox{SNIa}}}\r)_{\tiny{\hbox{screened}}} \,.
\ee

As for the Jordan frame in Eq.~\eqref{eq:JordanFrameScreenedDL}, we recover here the same gravitational luminosity distance as in standard GR and no difference to the electromagnetic luminosity distance.
This is in contrast to the difference found in the absence of screening of the local Planck mass evolution (Secs.~\ref{sec:JordanFrame} and \ref{sec:EinsteinFrame}) (for supernovae Type~Ia in case of powers $\gamma\neq1$ in the luminosity to Chandrasekhar mass relation) and implies that observationally compatible modifications of gravity, employing an efficient screening mechanism, do not leave an imprint in Standard Sirens.

\section{Implications for dark sector interactions and cosmic self-acceleration} \label{sec:darksectorinteractions}

At no point in the computations for the Standard Sirens test of gravity in Secs.~\ref{sec:JordanFrame}--\ref{sec:screeningmechanism} did we need to specify the coupling of the scalar field to dark matter particles.
Dark matter only entered the calculations indirectly such as through the Hubble friction causing a damping of the electromagnetic and gravitational waves.
Of relevance were instead the metrics to which baryons and photons couple minimally. We confirmed that the photons are unaffected by conformal transformations whereas we found that the baryons receive a shifted mass that impacts the Bohr radius and hence the  atomic emission lines (Sec.~\ref{sec:spinors}). As a consequence, we may allow for arbitrary couplings in the dark matter sector without impacting the Standard Sirens beyond the Hubble friction or secondary statistical effects from changing the cosmological structure that determines the paths of the electromagnetic and gravitational waves. Specifically, we shall consider here dark matter particles that follow geodesics of the metric $\check{g}_{\mu\nu}$ whereas baryonic fields and photons follow (see, e.g., Ref.~\cite{Gleyzes:2015pma})
\be
 \hat{g}_{\mu\nu} =C^{(\phi)}(\phi) \check{g}_{\mu\nu} +D^{(\phi)}(\phi) \p_\mu \phi \p_\nu \phi \,, \label{eq:confdisftrans}
\ee
where $\hat{g}_{\mu\nu}$ satisfies the GR field equations.
Hence, hats will refer to quantities in the Einstein frame.
Note that in addition to the conformal coupling studied in Secs.~\ref{sec:JordanFrame}--\ref{sec:screeningmechanism} we have also allowed for a disformal transformation, which in contrast does not preserve angles locally and affects the photons by changing their propagation speed in $\check{g}_{\mu\nu}$ to $\check{c}_\gamma = \sqrt{(C-D)/C}$, which follows from performing the disformal transformation on the electromagnetic action and examining the resulting equations of motion.

Because we have broken the assumption of universal coupling with baryonic and dark matter assumed to fall differently in a given gravitational field, we now may define two different Jordan frames: one where baryons and photons are minimally coupled to gravity (here coinciding with the Einstein frame) and one where dark matter is minimally coupled to gravity. To avoid confusion, when baryons and photons are minimally coupled, we shall refer to it as the \emph{baryon/photon frame}, and when dark matter is minimally coupled as the \emph{dark matter frame}.
In the baryon frame the photon and gravitational wave speeds are $\hat{c}_{\rm T} = \hat{c}_{\gamma}=c=1$ whereas in the dark matter frame these become $\cT= \check{c}_{\rm T}=\check{c}_{\gamma}$ \cite{Gleyzes:2015pma}.
The Horndeski action is invariant under transformations of the form of Eq.~\eqref{eq:confdisftrans}~\cite{Bettoni:2013diz}.
Schematically, in the baryon frame we have
\begin{align}
S[\hat{g}_{\mu\nu}, \phi, \psi_b, \psi_{dm}] \supset&  \int \dd^4 x \sqrt{-\hat{g}} \l(\frac{M_p^2}{2} \hat{R} + \mathcal{L}_b\l [\psi_b, \hat{g}_{\mu\nu} \r]\r) \\
&+ \int \dd^4 x \sqrt{-\hat{g}} \mathcal{L}_{dm}\l[ \psi_{dm},\frac{1}{C^{(\phi)}}\l(\hat{g}_{\mu\nu} - D^{(\phi)} \p_\mu\phi \p_\nu \phi \r) \r] \,,
\end{align}
where $\mathcal{L}_b\l [\psi_b, \hat{g}_{\mu\nu} \r]$ indicates that baryons are taken to be minimally coupled to $\hat{g}_{\mu\nu}$ whereas dark matter is nonminimally coupled to $\hat{g}_{\mu\nu}$.
Performing the transformation~\eqref{eq:confdisftrans} with free minimally coupled scalar field sector, one recovers the full Horndeski scalar-tensor theory with minimally coupled dark matter and nonminimally coupled baryons. 

For simplicity, we will work in the unitary gauge, in which the scalar field perturbations are absorbed in a change of the time coordinate such that $C= C(t)$ and $D= D(t) = D^{(\phi)}(t) \dot{\phi}^2$ (see, e.g., Ref.~\cite{Gleyzes:2015pma}).
Eq.~\eqref{eq:confdisftrans} simplifies to 
\be 
 \hat{g}_{\mu\nu} = C(\check{t}) \check{g}_{\mu \nu} + D(\check{t})\delta_\mu^0 \delta_\nu^0 \,.\label{UnitaryGaugeDisformal}
\ee
The invariant line element becomes
\begin{align}
\dd \hat{s}^2 &= \hat{g}_{\mu\nu} \dd \hat{x}^\mu \dd \hat{x}^\nu = (C \check{g}_{\mu\nu} +D\delta_\mu^0 \delta_\nu^0) \dd \hat{x}^\mu \dd \hat{x}^\nu = -(C- D)\dd \hat{t}^2 + C \hat{a}^2 \dd \mathbf{\hat{x}}^2 = - \dd \check{t}^2 + \check{a}^2 \dd \mathbf{\check{x}}^2 \,,
\end{align}
so that
\be
 \dd \check{t} \equiv \sqrt{C-D} \dd \hat{t} \,, \quad \dd \hat{\mathbf{x}} \equiv \dd \check{\mathbf{x}} \,, \quad \check{a} \equiv \sqrt{C} \hat{a} \,.
\ee

We will now determine the condition for cosmic acceleration to be fully attributed to the dark matter couplings $C$ and $D$ rather than arising from a dark energy contribution or a cosmological constant.
Such a self-acceleration effect is achieved when the scale factor in the baryon/photon frame, $\hat{a}$, is positively accelerated at late times but the acceleration of the scale factor in the dark matter frame, $\check{a}$, remains non-positive.
This is analogous to requiring for a genuine self-acceleration from modifying gravity that the scale factor at late times is positively accelerated in the Jordan frame but not so in the Einstein frame~\cite{Wang:2012kj,Lombriser:2015sxa}, which can be attributed to a significant evolution of a combination of $M(t)$ and $\cT(t)$~\cite{Lombriser:2015sxa}.
Note that this is equivalent to the recent distinction between force and fluid acceleration~\cite{Amendola:2019xqj} up to the subdominant dynamical contribution of the nonminimal couplings acting on the baryons and photons in the dark matter frame of a dark sector interaction model.
In the baryon frame, the observed late-time ($\hat{a}\gtrsim0.6$) accelerated expansion implies that
\be
 \frac{\dd^2\hat{a}}{\dd \hat{t}^2} = \hat{a}\hat{H}^2 \l(1+ \frac{\hat{H}'}{\hat{H}} \r) > 0 \,, \label{eq:JFacc}
\ee
where primes denote derivatives with respect to $\ln \hat{a}$ here and throughout this section. The scale factor in the dark matter frame must then instead satisfy (also see Refs.~\cite{Lombriser:2015sxa,Lombriser:2016yzn})
\begin{align}
\frac{\dd^2 \check{a}} {\dd \check{t}^2} 
= \frac{\hat{a} \hat{H}^2}{\sqrt{C-D}} \left( \mathcal{G}' + \left (1 + \frac{\hat{H}'}{\hat{H}} \right) \mathcal{G} \right) \leq 0 \,,
 \label{eq:inequality}
\end{align}
where we have defined 
\be 
\mathcal{G} \equiv  \frac{1}{\cT} \left( 1 + \frac{1}{2}\frac{C'}{C} \right) \,.
\ee
Recall that gravity is modified in this frame, and we have furthermore $C=M^2\cT$ with the evolving squared Planck mass $M^2$ \cite{Gleyzes:2015pma}.
Analogously to Ref.~\cite{Lombriser:2016yzn}, we can now solve the inequality~\eqref{eq:inequality}, setting $C$ to unity at the onset of cosmic acceleration $\hat{t}_{acc}$ in Eq.~\eqref{eq:JFacc} and requiring $C'(\hat{t}_{acc})/C(\hat{t}_{acc})=0$ and $D(\hat{t}_{acc})=0$, to find
\be \label{eq:MinimalPlanckMass}
 C(t)= M^2(t) \cT(t) \leq \left( \frac{a_{acc}}{a(t)}\right)^2 \exp \left( K_1 \int_{t_{acc}}^t \frac{\cT(t') \dd t'}{ a(t')} \right) \,,
\ee
where $a_{acc} \equiv a(t_{acc})$ denotes the scale factor at the onset of cosmic acceleration and if adopting a $\Lambda$CDM expansion history,
\be
K_1 \equiv 2H_0 a_{acc}\sqrt{3(1+\Omega_m)} \,, \quad a_{acc} = \left( \frac{\Omega_m}{2 (1 - \Omega_m)}\right)^{1/3} \,.
\ee
For $\cT=1$, we recover the inequality condition for self-acceleration found in Ref.~\cite{Lombriser:2016yzn}.
This also recovers the significant running in $C'/C = (M^2)'/M^2 + \cT'/\cT$ required for self-acceleration in Ref.~\cite{Lombriser:2015sxa}, however with a different power of $\cT$, which stems from the adoption of the Einstein-Friedmann frame in Ref.~\cite{Lombriser:2015sxa} rather than the dark matter frame.

Note that since we are in the dark matter frame, this would be the resulting effective expansion of a modified theory of gravity where the dark matter is minimally coupled but the baryons are nonminimally coupled in such a way that a transformation into Einstein frame would make baryons minimally coupled. The effective total matter energy density parameter $\Omega_m$ hence needs to be interpreted accordingly. In the baryon frame, Eq.~\eqref{eq:MinimalPlanckMass} represents a minimal criteria for cosmic acceleration to be caused by dark sector interactions specified by $C$ and $D$.
The criteria is minimal, or conservative, in the sense that some interaction may already be required to allow the universe to reach the steady state $\dd^2\check{a}/\dd\check{t}^2=0$.

Importantly, for nonuniversal couplings $C$ and $D$, the dark degeneracy between the two in the large-scale structure is restored~\cite{Lombriser:2014ira,Lombriser:2015sxa}.
This is because $\check{c}_T=\check{c}_\gamma$ is no longer constrained by the arrival time difference of gravitational waves and light and hence does not break the degeneracy. In other words, the speed of gravitational waves and light is the same in the dark matter frame and in the baryon frame independently of the values of $C$ and $D$.
Hence, in contrast to modified gravity~\cite{Lombriser:2016yzn}, an observationally compatible self-acceleration effect emanating from the couplings $C$ and $D$ can be realised as long as baryons and photons remain uncoupled in the baryon frame, which is evident as the couplings and scalar field sector can be chosen to mimic standard cosmology both in the cosmological background evolution and the large-scale structure~\cite{Lombriser:2015sxa}.
We leave an analysis of such scenarios to future work.

\section{Conclusions} \label{sec:conclusions}

Standard Sirens tests of gravity have been proposed as a powerful test of gravity that exploits the difference in electromagnetic and gravitational wave luminosity distances arsing from the cosmological time variation of the effective Newton constant in modified gravity theories.
The role of screening mechanisms that must be employed to comply with stringent astrophysical tests of GR, however, has so far remained unclear.

We analysed here in detail the workings of the Standard Sirens test for Horndeski scalar-tensor theories abiding to a luminal speed of gravity, where a particular emphasis was put on the impact of screening mechanisms.
We find that screening reduces the gravitational wave luminosity distance to its GR expression, thus leaving no observable signature for viable cosmological gravitational modifications in Standard Sirens.
Previously it was assumed that the sources of electromagnetic and gravitational radiation should be screened, causing GR-like emissions, but that the gravitational wave luminosity distance would be modified by the cosmological propagation due to the evolving Planck mass in the background.
We found here that this is not the case.

In our analysis, we have first considered a uniform time variation of the effective Planck mass in the Jordan frame in the absence of its local screening at the source and at the observer, which may also encompass the Vainshtein mechanism. We derived the gravitational wave and electromagnetic luminosity distances from the action for a flat FLRW universe, discussing in detail the propagation effects for gravitational waves and for electromagnetic radiation.
While the gravitational wave amplitudes are damped differently to GR, the production of the waves from the inspiral phase of compact binaries is not affected by the cosmological time evolution of the gravitational coupling and hence leads to a GR-like emission even in the absence of screening.
This can be attributed to a degeneracy of the chirp mass with the redshift of the source.
In contrast, electromagnetic emissions at the source can be modified. The production of supernovae Type~Ia, for example, is generally modified in the presence of an evolving Planck mass, due to the dependence of the Chandrasekhar mass on the gravitational coupling, changing the electromagnetic luminosity distances inferred with respect to GR. In particular, depending on the relation between the peak luminosity of supernovae to the Chandrasekhar mass, the change of the electromagnetic luminosity distance may become equivalent to that for gravitational waves, leaving no signature in Standard Sirens even in the absence of screening.

We performed these computations also in the Einstein frame, which involved some subtleties such as geodesic motion and the correct, frame-independent notion of redshift.
We confirmed that our results for the observable quantities remain invariant under the frame transformation.

Next, we implemented a generalised screening mechanism that suppresses the time variation of the effective gravitational coupling locally.
We found that in the approximation where the scalar field amplitude varies on scales much larger than the gravitational wavelength, the gravitational wave luminosity distance reduces to that of GR.
Likewise, GR is also restored in the production of supernovae Type~Ia, leaving electromagnetic luminosity distances unaffected.
Hence, in the presence of screening mechanisms as required for viable gravitational modifications on cosmological scales, the Standard Sirens and the supernovae Type Ia of the modified theory become indistinguishable to their GR counterparts.

Our results have important implications for Standard Sirens tests of gravity. Either the gravitational modifications of consideration do not employ a screening mechanism of the local Planck mass evolution, in which case these must comply with stringent astrophysical constraints such as inferred from lunar laser ranging, lying several orders of magnitude beyond the sensitivity of Standard Sirens test and leaving no detectable signature for those -- there may even be no signature at all in case of using supernovae Type Ia and a linear relationship of the peak luminosity to the Chandrasekhar mass. Or the models are equipped with a screening mechanism, in which case the luminosity distances become indistinguishable, as in GR.
Modified gravity effects will then be limited to a potentially different cosmological background expansion history and secondary effects in the propagation of light and gravitational waves through modified cosmological structures.

Despite this limiting result, it should be stressed that Standard Sirens provide valuable independent tests of gravity, which differ fundamentally from the more stringent astrophysical probes, a circumstance that may generally be important for more exotic gravitational modifications and related scenarios.
For example, while our result likely applies to more general models for which the extra friction in the gravitational wave propagation equation arises from a time variation of the effective Planck mass, being the case for many interesting models, it does not hold for some non-local gravity or extra-dimensional models.

Finally, using our results from the Einstein frame, we have examined dark sector interactions, where baryons and photons remain minimally coupled whereas dark matter couples nonminimally to a disformally related metric.
We showed that the gravitational wave propagation is not modified for conformal and disformal dark sector interactions, which follows from a generalisation of our results obtained for modified gravity.
As a consequence of that, these interactions can yield an observationally compatible self-acceleration effect emanating from the dark sector couplings by restoration of a dark degeneracy that is broken in modified gravity due to the luminal speed of gravity.
We provide a minimal criteria for the conformal and disformal couplings to dark matter to source cosmic self-acceleration. A detailed exploration of these models is left to future work.

\section*{Acknowledgements}
The authors thank Camille Bonvin, Ruth Durrer, Pierre Fleury, and Michele Maggiore for useful discussions.
C.D.~and L.L.~were supported by a Swiss National Science Foundation (SNSF) Professorship grant (No.~170547).
Please contact the authors for access to research materials.

\bibliographystyle{JHEP}
\bibliography{references}

\end{document}